\documentclass{llncs}

\usepackage{mathptmx}
\usepackage{helvet}
\usepackage{courier}
\usepackage{multicol}
\usepackage[bottom]{footmisc}
\usepackage{amssymb}
\usepackage{rotating}

\usepackage{epsfig}
\usepackage{latexsym}
\usepackage{graphicx}
\usepackage{color}
\usepackage{url}
\usepackage{amsfonts}
\usepackage{pgf,pgfarrows,pgfnodes,pgfautomata,pgfheaps,pgfshade}
\usepackage{tikz}

\usetikzlibrary{arrows,decorations.pathmorphing,backgrounds,positioning,fit,petri}
\usepackage{etex}

\usepackage{stmaryrd}
\usepackage{amsmath}

\usepackage{bussproofs}

\usepackage{lineno}

\newcommand{\post}[1]{\mbox{$#1^{\bullet}$}}
\newcommand{\pre}[1]{\mbox{$^{\bullet}#1$}}

\newcommand{\fine}{{\mbox{ }\nolinebreak\hfill{$\Box$}}}

\newcommand{\deriv}[1]{{\mbox{${\:\stackrel{#1}{\longrightarrow}\:}$}}}

\newcommand{\nderiv}[1]{\nrightarrow}

\newcommand{\eqdef}{ \doteq }

\newcommand{\bigfrac}[2]{
\renewcommand{\arraystretch}{1.5}
\begin{array}{c}#1\\
\hline
#2
\end{array}}
\newcommand{\restr}[1]{\mbox{$({\bf\nu} #1)$}}
\newcommand{\para}{\mbox{$\,|\,$}}

\newcommand{\1}{\mbox{\bf 1}}

\renewcommand{\mid}{\;\;\big|\;\;}

\newcommand{\nat}{{\mathbb N}}

\newcommand{\atom}[1]{\mbox{$\lessdot #1 \gtrdot$}}

\begin{document}

 \pagestyle{headings}

\title{Decidable Reversible Equivalences for Finite Petri Nets
}
\author{Roberto Gorrieri \and Ivan Lanese\\
\institute{Universit\`a di Bologna, Dipartimento di Informatica --- Scienza e Ingegneria\\
Mura A. Zamboni, 7,
40127 Bologna, Italy}
\email{{\small roberto.gorrieri@unibo.it \; ivan.lanese@unibo.it}}
}

\maketitle

\begin{abstract}
In the setting of Petri nets, we prove that {\em causal-net bisimilarity} \cite{G15,Gor22,Gor25a}, which is a 
refinement of history-preserving bisimilarity \cite{RT88,vGG89,DDM89},
and the novel {\em hereditary} causal-net bisimilarity, which is a refinement of 
hereditary history-preserving bisimilarity \cite{Bed91,JNW96}, do coincide.
This means that causal-net bisimilarity is a {\em reversible behavioral equivalence}, 
as causal-net bisimilar markings 
not only are able to match each other's forward
transitions, but also backward transitions by undoing performed events.
Causal-net bisimilarity can be equivalently formulated as 
{\em structure-preserving bisimilarity} \cite{G15,Gor25a}, that is decidable on finite 
bounded Petri nets \cite{CG21a}.
Moreover, place bisimilarity \cite{ABS91}, that we prove to be finer than causal-net 
bisimilarity, is also reversible and it was proved 
decidable for finite Petri nets in \cite{Gor21decid,Gor25a}.
These results offer two decidable reversible behavioral equivalences in the true concurrency 
spectrum, which are alternative to the coarser
hereditary history-preserving bisimilarity \cite{Bed91,JNW96},
that, unfortunately, is undecidable even for safe Petri nets \cite{JNS03}.
\end{abstract}

%
\section{Introduction}
%

Traditionally, a behavioral equivalence relates states (or markings, or configurations) that have equivalent future behavior.
For instance, in a Deterministic Finite Automaton (DFA) \cite{HMU01}, two states are equivalent if they recognize the same language 
(i.e., if their forward computations, 
leading to a final state, give rise to the same words). Similarly, in a labeled transition system (LTS) \cite{Kel76,GV15},
two states are bisimulation equivalent \cite{Park81,Mil89} if they simulate each 
other's behavior in the direction of the arrows, leading to pairs of bisimilar states.

When considering systems that should be able to undo past events to, e.g.,
take care of missed dependencies in optimistic simulation~\cite{CPF99}, redo actions in case of errors in industrial robots~\cite{LSE15}, or find by going backwards which bug caused a visible misbehavior in reversible debugging~\cite{MMS17,HL+20}, 
it is useful that the behavioral equivalence of interest considers also backward transitions, related to the past history.
We will call {\em reversible behavioral equivalence} a behavioral equivalence relation 
such that related systems are able to match not only their forward behavior, but also their past behavior, so that, 
by undoing some performed events, the reached states are 
still equivalent.

In their seminal paper, De Nicola, Montanari and Vaandrager \cite{DMV90} proposed 
{\em back and forth} bisimulation, 
as an enhancement of ordinary bisimulation on LTSs \cite{Mil89,GV15}, and proved 
that these two definitions define the same equivalence
relation. Similarly, they also proved that, when considering LTSs labeled also with the invisible 
action $\tau$, the back and forth
version of branching bisimilarity \cite{vGW96,GV15} turns out to coincide with branching bisimilarity itself.
We will call  {\em self-reversible} a behavioral equivalence which is a `forward' equivalence relation, i.e., a relation defined
on the forward behavior only, that, nonetheless, is reversible.
Hence, according to \cite{DMV90}, on LTSs both bisimulation equivalence \cite{Mil89} 
and branching bisimulation equivalence \cite{vGW96} are self-reversible behavioral relations.

On LTSs we have to be a bit careful about the actual meaning of the past behavior. 
As observed in \cite{DMV90}, the past history is the
actual computation that starts from the initial state and reaches the current state. 
Only these transitions are to be considered in the
back (and forth) bisimulation game, as these are the only ones that can be undone. 
This view is correct under the assumption that the LTS is modeling a 
{\em sequential system}, so that it cannot perform concurrent events.

If we want to model a concurrent system, it is better to use some truly concurrent model of concurrency, 
such as {\em event structures} \cite{Win87}, {\em configuration structures} \cite{GP09} or {\em Petri nets} \cite{Pet81,Gor17,Gor25a}.
If we consider configuration structures, the past history is not the actual computation path, 
obtained by performing an event sequence $\sigma$, from the initial (empty)
configuration to the current configuration, but rather the whole set of configurations that can be reached from the 
initial configuration by performing any permutation of $\sigma$ that is compatible with the partial order that $\sigma$ induces. 
For instance, if $X_0 \deriv{e_1} X_1 \deriv{e_2} X_2$, where $e_1$ and $e_2$ are concurrent events, then
we have also to check the configuration $X_1'$ such that $X_0 \deriv{e_2} X_1' \deriv{e_1} X_2$, because this is reachable from
$X_2$ by undoing the event $e_1$.
This notion of reversibility takes the name of \emph{causal-consistent reversibility} and it has been first introduced in~\cite{DK04} in the 
context of reversible CCS. Since then, it has been applied to a number of models and languages, including
the $\pi$-calculus~\cite{CKV13}, the higher-order $\pi$-calculus~\cite{LMS16}, Place-Transition Petri nets~\cite{MMU20}, 
event structures~\cite{PU07}, and the Erlang programming language~\cite{LNPV18}. See~\cite{G+23} for a recent survey.

On truly concurrent models of concurrency, a popular behavioral equivalence is {\em history-preserving
bisimilarity} \cite{RT88,vGG89,DDM89} (hp-bisimilarity, for short), that on Petri nets takes the form of
{\em fully concurrrent bisimilarity} \cite{BDKP91}. A back and forth version of hp-bisimilarity was proposed 
in \cite{Bed91,JNW96} under the name of 
{\em hereditary history-preserving bisimilarity}  (hhp-bisimilarity, for short), that enhances hp-bisimilarity by requiring 
that the hp-bisimulation game between two systems
also holds when going backward in their history. Hence, hhp-bisimilarity is a reversible equivalence, 
actually the coarsest reversible bisimulation-based, behavioral relation
fully respecting causality.
This is also in line with the results in~\cite{AC20}, which prove hhp-bisimilarity equivalent to a back and forth bisimilarity in the context of 
reversible CCS.
However, hp-bisimilarity is not self-reversible because
hp-bisimilarity is coarser than hhp-bisimilarity.
Moreover, while hp-bisimilarity is decidable on BBP nets \cite{FJSL10,Gor22}, safe finite Petri nets \cite{Vog91} and 
on bounded finite Petri nets \cite{MP97,CG21a}, 
unfortunately hhp-bisimilarity has much poorer decidability results; in fact,
it is decidable on BBP nets \cite{FJSL10},
but
it is undecidable even on safe nets \cite{JNS03}, hence making its usability for practical problems very limited.

The main aim of this paper is to find {\em reversible}, {\em decidable}, {\em finer} approximations 
of hereditary history-preserving bisimilarity over finite Petri nets.

About {\em coarser} decidable approximations, there is an interesting result by Fr\"{o}schle and Hildebrandt \cite{FH99}, who 
have proposed an infinite hierarchy of bisimilarity notions, all decidable on safe finite Petri nets, that refine hp-bisimilarity, 
and are coarser than hhp-bisimilarity, such that hhp-bisimilarity is the intersection of all these bisimilarities in the hierarchy. 
However, this result is not very useful in practice because coarser approximations of hhp-bisimilarity do not ensure reversibility.

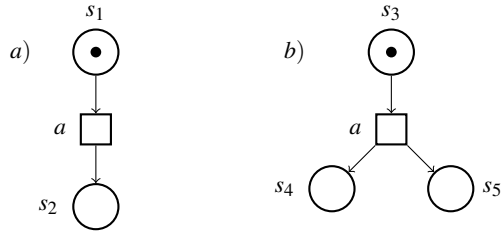
\begin{figure}[t]
\centering
\begin{tikzpicture}[
every place/.style={draw,thick,inner sep=0pt,minimum size=6mm},
every transition/.style={draw,thick,inner sep=0pt,minimum size=4mm},
bend angle=30,
pre/.style={<-,shorten <=1pt,>=stealth,semithick},
post/.style={->,shorten >=1pt,>=stealth,semithick}
]
\def\eofigdist{3.3cm}
\def\eodist{0.5cm}
\def\eodisty{0.9cm}

\node (a) [label=left:$a)\qquad $]{};

\node (p1) [place, tokens=1]  [label=above:$s_1$] {};
\node (t1) [transition] [below =\eodist of p1,label=left:$a\;$] {};
\node (p2) [place] [below =\eodist of t1,label=left:$s_2\;$] {};

\draw  [->] (p1) to (t1);
\draw  [->] (t1) to (p2);

  \node (b) [right={3.1cm} of a,label=left:$b)\quad$] {};

\node (p3) [place,tokens=1]  [right=\eofigdist of p1, label=above:$s_3$] {};
\node (t3) [transition] [below =\eodist of p3,label=left:$a\;$] {};
\node (p4) [place]  [below left =\eodist of t3, label=left:$s_4\;$] {};
\node (p5) [place] [below right=\eodist of t3,label=right:$s_5$] {};

\draw  [->] (p3) to (t3);
\draw  [->] (t3) to (p4);
\draw  [->] (t3) to (p5);

\end{tikzpicture}
\caption{Two fully-concurrent bisimilar markings, which are not causal-net bisimilar}
\label{net-fc-place}
\end{figure}

A behavioral equivalence finer than fully concurrent bisimilarity \cite{BDKP91} on 
Petri nets (i.e., finer than hp-bisimilarity) is 
{\em causal-net bisimilarity} \cite{G15,Gor22,Gor25a}, denoted by $\sim_{cn}$, 
that relates markings generating net processes that share the same underlying {\em causal net} (also called 
{\em occurrence net}) \cite{GR83,BD87,Old,G15,Gor22}. 
Hence, causal-net bisimilar markings must have the same size and the matching transitions must have the same shape.
As a simple example, consider the nets in Figure \ref{net-fc-place}, where
$s_1$ and $s_3$ are fully concurrent bisimilar (i.e., hp-bisimilar), because both can perform just one $a$-labeled event.
However, it is not possible to build a causal-net bisimulation relating $s_1$ and $s_3$  
because they generate different causal nets (as the post-sets of the
two transitions have different size).

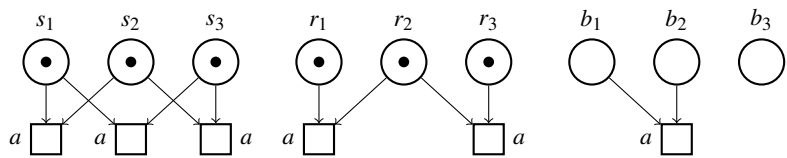
\begin{figure}[t]
\centering

\begin{tikzpicture}[
every place/.style={draw,thick,inner sep=0pt,minimum size=6mm},
every transition/.style={draw,thick,inner sep=0pt,minimum size=4mm},
bend angle=42,
pre/.style={<-,shorten <=1pt,>=stealth,semithick},
post/.style={->,shorten >=1pt,>=stealth,semithick}
]
\def\eofigdist{3cm}
\def\eodist{0.5cm}
\def\eodisty{1.5cm}

\node (p1) [place,tokens=1]  [label=above:$s_1$] {};
\node (p2) [place,tokens=1]  [right=\eodist of p1,label=above:$s_2$] {};
\node (p3) [place,tokens=1] [right=\eodist of p2,label=above:$s_3$] {};
\node (t1) [transition] [below=\eodist of p1,label=left:$a$] {};
\node (t2) [transition] [below=\eodist of p2,label=left:$a$] {};
\node (t3) [transition] [below=\eodist of p3,label=right:$a$] {};

\draw  [->] (p1) to (t1);
\draw  [->] (p2) to (t1);
\draw  [->] (p1) to (t2);
\draw  [->] (p3) to (t2);
\draw  [->] (p2) to (t3);
\draw  [->] (p3) to (t3);

\node (q1) [place,tokens=1]  [right=\eofigdist of p1,label=above:$r_1$] {};
\node (q2) [place,tokens=1]  [right=\eodist of q1,label=above:$r_2$] {};
\node (q3) [place,tokens=1]  [right=\eodist of q2,label=above:$r_3$] {};
\node (s1) [transition] [below=\eodist of q1,label=left:$a$] {};
\node (s2) [transition] [below=\eodist of q3,label=right:$a$] {};

\draw  [->] (q1) to (s1);
\draw  [->] (q2) to (s1);
\draw  [->] (q2) to (s2);
\draw  [->] (q3) to (s2);

\node (b1) [place]  [right=\eofigdist of q1,label=above:$b_1$] {};
\node (b2) [place]  [right=\eodist of b1,label=above:$b_2$] {};
\node (b3) [place]  [right=\eodist of b2,label=above:$b_3$] {};
\node (u1) [transition] [below=\eodist of b2,label=left:$a$] {};

\draw  [->] (b1) to (u1);
\draw  [->] (b2) to (u1);

\end{tikzpicture}
\caption{Two further fully-concurrent bisimilar markings,
  which are
  not causal-net bisimilar}
\label{cn-vs-icn-fig}
\end{figure}

Even more interesting are the nets in Figure \ref{cn-vs-icn-fig}, that reveal a more subtle property of causal-net bisimilarity.
Consider the two markings $s_1 \oplus s_2 \oplus s_3$ and $r_1 \oplus r_2 \oplus r_3$.
It is easy to see that these two markings are fully concurrent bisimilar, because what they can do is just one $a$-labeled event.
Moreover, they generate the same causal nets, all isomorphic to the causal net on the right of Figure \ref{cn-vs-icn-fig}.
Nonetheless, these two markings are not causal-net bisimilar.
In fact, a causal-net bisimulation relating these two markings 
should be a relation composed of triples of the form $(\rho_1, C, \rho_2)$ where $(C, \rho_1)$ is a process for $s_1 \oplus s_2 \oplus s_3$,
while $(C, \rho_2)$ is a process for $r_1 \oplus r_2 \oplus r_3$.
Whatever are the initial {\em foldings} $\rho_1^0$ and $\rho_2^0$ from the initial causal net, composed of the three conditions 
$b_1, b_2, b_3$ only,
to places of the two nets, it is always 
possible for the first net to 
perform a transition that is not matched by the second net.
For instance, assume that these mappings are the trivial ones, i.e., $\rho_1^0$ maps $b_i$ to $s_i$ and $\rho_2^0$ maps $b_i$ 
to $r_i$ for $i = 1, 2, 3$. Then, if the first net performs
the transition $(s_1 \oplus s_3, a, \theta)$, the second net cannot reply, because $r_1 \oplus r_3$ is stuck.

In fact, as observed in \cite{CG21a}, in a fully concurrent bisimulation, it is not necessary to fix in advance
the definition of the foldings from the maximal places of the causal nets to the current markings 
{\em before} the bisimulation game starts; on the contrary, they can be
actually constructed progressively as long as the computation proceeds. 
This property, i.e., fixing the folding before the bisimulation game starts, is also at the base of an important property
of causal-net bisimilarity; in fact, as observed in \cite{G15}, it is the coarsest equivalence that respects 
{\em inevitability} \cite{MOP89}, i.e.,  if two
markings are causal-net bisimilar, and in one the occurrence of a certain action is inevitable, then so it is in the other one. 
Moreover, causal-net bisimilarity can be characterized in a very simple manner over 
Petri nets as {\em structure-preserving bisimilarity} \cite{G15,Gor25a}, that is decidable on bounded, finite Petri nets \cite{CG21a}.

Another interesting behavioral equivalence over Petri nets is \emph{place bisimilarity} \cite{ABS91,Gor25a}, that is  finer than
causal-net bisimilarity, but decidable on the whole class of finite Petri nets \cite{Gor21decid,Gor25a}.
Informally, a binary relation $R$ over the set $S$ of places of a Petri net  is a place bisimulation if for all 
markings $m_1$ and $m_2$
which are {\em bijectively} related via $R$ (denoted by $(m_1, m_2) \in R^\oplus$, where $R^\oplus$ is called the 
{\em additive closure} of $R$), if $m_1$ can perform transition $t_1$, reaching marking $m_1'$,
then $m_2$ can perform a transition $t_2$, reaching marking $m_2'$, 
such that the pre-sets of $t_1$ and $t_2$ are related by $R^\oplus$ (i.e., $(\pre{t_1}, \pre{t_2}) \in R^\oplus)$,
the label of $t_1$ and $t_2$ is the same, the post-sets of $t_1$ and $t_2$ are related by $R^\oplus$ 
(i.e., $(\post{t_1}, \post{t_2}) \in R^\oplus)$, and 
also for the reached markings we have $(m_1', m_2') \in R^\oplus$;
and symmetrically if $m_2$ moves first.  
Two markings $m_1$ and $m_2$ are place bisimilar, denoted by $m_1 \sim_p m_2$, if  a place bisimulation $R$ exists such
that $(m_1, m_2) \in R^\oplus$.  

The main contributions of this paper can be summarized as follows:
\begin{itemize}
\item The definition of {\em hereditary causal-net bisimulation}, by extending the original definition 
with backward transitions, whose induced 
equivalence is denoted by $\sim_{hcn}$.
\item The proof that $\sim_{cn}$ and $\sim_{hcn}$ coincide; this means that causal-net bisimilarity is self-reversible. 
\item Analogously, we also define {\em hereditary fully concurrent bisimulation}, 
by extending the original definition in \cite{BDKP91} 
with backward transitions; the induced reversible behavioral relation, which is the porting on Petri nets
of hereditary history-preserving bisimilarity, is denoted by $\sim_{hfc}$.
\item The observation that $\sim_{cn}$ is finer than $\sim_{hfc}$.
\item The proof that place bisimilarity $\sim_p$ is finer than
causal-net bisimilarity $\sim_{cn}$.
\item The observation that place bisimilarity $\sim_p$, being finer than $\sim_{cn}$,
is also self-reversible. In fact, we show that place bisimilar markings are able not only to match their future behavior, 
but also to match their past behavior. Since place bisimilarity is decidable 
on finite Petri nets, it can be used as an 
alternative to causal-net bisimilarity (or, better, the equivalent structure-preserving bisimilarity \cite{G15}) 
when the marked Petri net is unbounded.
\item A small case study, modeling two unbounded producer--consumer systems, that illustrates the real applicability of the proposed reversible equivalences.
\end{itemize} 

The paper is organized as follows. Section \ref{def-sec} recalls the basic definitions about Petri nets, 
including place bisimulation and structure-preserving bisimulation.
Section \ref{causal-petri-sec} defines the causal semantics of Petri nets, including
 causal-net bisimulation,  fully concurrent bisimulation,
and the novel definitions of   {\em hereditary causal-net bisimulation} and {\em hereditary fully concurrent bisimulation};
this section also contains the proofs that causal-net bisimilarity and hereditary causal-net bisimilarity coincide and that
place bisimilarity implies causal-net bisimilarity.
Section \ref{place-reverse-sec} argues that also place bisimilarity is self-reversible.
Section \ref{case-sec} discusses a small case study about two producer--consumer systems, showing the real applicability of the approach.
Finally, Section \ref{conc-sec} overviews related literature, formulates a conjecture about the coarsest self-reversible 
causality-respecting bisimulation-based behavioral relation and hints at some possible
future research.

%
\section{Petri Nets} \label{def-sec}
%

\begin{definition}\label{multiset}{\bf (Multiset)}\index{Multiset}
Let $\nat$ be the set of natural numbers. 
A {\em multiset} over a finite set $S$ is a function $m: S \rightarrow\nat$. 
The {\em support set} $dom(m)$ of $m$ is $\{ s \in S \mid m(s) \neq 0\}$. 
The set of all multisets 
over $S$,  denoted by ${\mathcal M}(S)$, is ranged over by $m$. 
We write $s \in m$ if $m(s)>0$.
The {\em multiplicity} of $s$ in $m$ is given by the number $m(s)$. The {\em size} of $m$, denoted by $|m|$,
is the number $\sum_{s\in S} m(s)$, i.e., the total number of its elements.
A multiset $m$ such 
that $dom(m) = \emptyset$ is called {\em empty} and is denoted by $\theta$.
We write $m \subseteq m'$ if $m(s) \leq m'(s)$ for all $s \in S$. 
{\em Multiset union} $\_ \oplus \_$ is defined as follows: $(m \oplus m')(s)$ $ = m(s) + m'(s)$; it 
is commutative, associative and has $\theta$ as neutral element. 
{\em Multiset difference} $\_ \ominus \_$ is defined as 
$(m_1 \ominus m_2)(s) = max\{m_1(s) - m_2(s), 0\}$.
The {\em scalar product} of a number $j$ with $m$ is the multiset $j \cdot m$ defined as
$(j \cdot m)(s) = j \cdot (m(s))$. By $s_i$ we also denote the multiset with $s_i$ as its only element.
Hence, a multiset $m$ over $S = \{s_1, \ldots, s_n\}$
can be represented as $k_1\cdot s_{1} \oplus k_2 \cdot s_{2} \oplus \ldots \oplus k_n \cdot s_{n}$,
where $k_j = m(s_{j}) \geq 0$ for $j= 1, \ldots, n$.
\fine
\end{definition}

\begin{definition}\label{pt-net-def}{\bf (Place/Transition Petri net)}
A labeled {\em Place/Transition} Petri net (P/T net, for short) is a tuple $N = (S, A, T)$, where
\begin{itemize}
\item 
$S$ is the finite set of {\em places}, ranged over by $s$ (possibly indexed),
\item 
$A$ is the finite set of {\em labels}, ranged over by $a, b, c, ...$ (possibly indexed), and
\item 
$T \subseteq {(\mathcal M}(S) \setminus \{\theta\}) \times A \times {\mathcal M}(S)$ 
is the finite set of {\em transitions}, 
ranged over by $t$ (possibly indexed).
\end{itemize}

Given a transition $t = (m, a, m')$,
we use the notation $\pre t$ to denote its {\em pre-set} $m$ (which cannot be an empty multiset) of tokens to be consumed, $\ell(t)$ for its {\em label} $a$, and $\post t$ to denote its {\em post-set} $m'$ of tokens to be produced.
We define the  function
$\mathsf{flow}: (S \times T) \cup (T \times S) \rightarrow \nat$ as follows:
for all $s \in S$, for all $t \in T$,
$\mathsf{flow}(s,t) = \pre{t}(s)$ and $\mathsf{flow}(t,s) = \post{t}(s)$
(note that $\pre{t}(s)$ and $\post{t}(s)$ are integers, representing the multiplicity of $s$ in $\pre{t}$ and $\post{t}$, respectively).
We denote by $F$ the {\em flow relation} 
$\{(x,y) \mid x,y \in S \cup T \, \wedge \, \mathsf{flow}(x,y) > 0\}$.
Finally, we 
define pre-sets and post-sets for places as follows: $\pre s = \{t \in T \mid s \in \post t\}$
and $\post s = \{t \in T \mid s \in \pre t\}$. 
\fine
\end{definition}

Graphically, a place is represented by a little circle and a transition by a little box. These are 
connected by directed arcs, which
may be labeled by a positive integer, called the {\em weight}, to denote the number of tokens 
consumed (when the arc goes from a place to the transition) or produced (when the arc goes form the transition to a
place) by the execution of the transition;
if the number is omitted, then the weight default value of the arc is $1$.

\begin{definition}\label{net-system}{\bf (Marking, P/T net system)}
Given a P/T Petri net $N = (S, A, T)$, a multiset over $S$  is called a {\em marking}. Given a marking $m$ and a place $s$, 
we say that the place $s$ contains $m(s)$ {\em tokens}, graphically represented by $m(s)$ bullets
inside place $s$.
A {\em P/T net system} $N(m_0)$ is a tuple $(S, A, T, m_{0})$, where $(S,A, T)$ is a P/T net and $m_{0}$ is  
a marking over $S$, called
the {\em initial marking}. We also say that $N(m_0)$ is a {\em marked} net.
\fine
\end{definition}

\begin{definition}\label{firing-system}{\bf (Enabling, firing sequence, reachable marking, safe net)}
Given a P/T Petri net $N = (S, A, T)$, a transition $t $ is {\em enabled} at a marking $m$, 
denoted by $m[t\rangle$, if $\pre t \subseteq m$. 
The execution (or {\em firing}) of  $t$, enabled at $m$, produces the marking $m' = (m \ominus  \pre t) \oplus \post t$. 
This is written $m[t\rangle m'$. 
A {\em firing sequence} starting at $m$ is defined inductively as follows:
\begin{itemize}
\item $m[\epsilon\rangle m$ is a firing sequence (where $\epsilon$ denotes an empty sequence of transitions) and
\item if $m[\sigma\rangle m'$ is a firing sequence and $m' [t\rangle m''$, then
$m [\sigma t\rangle m''$ is a firing sequence. 
\end{itemize}
The set of {\em reachable markings} from $m$ is 
$[m\rangle = \{m' \mid  \exists \sigma.$ $
m[\sigma\rangle m'\}$. 
The P/T net system $N = $ $(S, A, T, m_0)$ is  {\em safe} if  for each 
$m \in [m_0\rangle$ and for all $s \in S$, we have $m(s) \leq 1$.
\fine
\end{definition}

Note that the reachable markings of a P/T net can be countably infinitely many when the net is not bounded, i.e.,
when the number of tokens on some places can grow unboundedly.

Now we recall a simple behavioral equivalence on P/T nets, defined directly over the markings of the net, which 
compares two markings with respect to their sequential behavior.

\begin{definition}\label{def-int-bis}{\bf (Interleaving Bisimulation)}
Let $N = (S, A, T)$ be a P/T net. 
An {\em interleaving bisimulation} is a relation
$R\subseteq {\mathcal M}(S) \times {\mathcal M}(S)$ such that if $(m_1, m_2) \in R$
then
\begin{itemize}
\item $\forall t_1$ such that  $m_1[t_1\rangle m'_1$, $\exists t_2$ such that $m_2[t_2\rangle m'_2$ 
with $\ell(t_1) = \ell(t_2)$ and $(m'_1, m'_2) \in R$,
\item $\forall t_2$ such that  $m_2[t_2\rangle m'_2$, $\exists t_1$ such that $m_1[t_1\rangle m'_1$ 
with $\ell(t_1) = \ell(t_2)$ and $(m'_1, m'_2) \in R$.
\end{itemize}
Two markings $m_1$ and $m_2$ are {\em interleaving bisimilar}, 
denoted by $m_1 \sim_{int} m_2$, if there exists an interleaving bisimulation $R$ such that $(m_1, m_2) \in R$.
\fine
\end{definition}

Interleaving bisimilarity was proved undecidable in \cite{Jan95} for P/T nets having at least two unbounded places, 
with a proof based on the comparison of two {\em sequential} P/T nets, 
where a P/T net is sequential if it does not offer any concurrent behavior. Hence, interleaving bisimulation equivalence is 
undecidable even for the subclass of sequential finite P/T nets. Esparza observed in \cite{Esp98} that over sequential P/T nets
all the non-interleaving 
bisimulation-based equivalences (in the spectrum ranging from interleaving bisimilarity to fully-concurrent bisimilarity \cite{BDKP91})
collapse to interleaving bisimilarity.
Hence, the proof in \cite{Jan95} applies to all these
non-interleaving bisimulation equivalences as well.

\subsection{Place Bisimulation}\label{place-bis-sec}

We now present the place bisimulation idea, introduced originally in \cite{ABS91}.
This proposal is an 
improvement of a behavioral relation proposed earlier by Olderog in \cite{Old} on safe 
nets (called {\em strong bisimulation}), which was not transitive. 
First, an auxiliary definition.

\begin{definition}\label{add-eq}{\bf (Additive closure)}
Given a P/T net $N = (S, A, T)$ and a {\em place relation} $R \subseteq S \times S$, we define a {\em marking relation}
$R^\oplus \, \subseteq \, {\mathcal M}(S) \times {\mathcal M}(S)$, called 
the {\em additive closure} of $R$,
as the least relation induced by the following axiom and rule.\\

$\begin{array}{lllllllllll}
 \bigfrac{}{(\theta, \theta) \in  R^\oplus} & \;   &   \; 
 \bigfrac{(s_1, s_2) \in R \quad  (m_1, m_2) \in R^\oplus }{(s_1 \oplus m_1, s_2 \oplus m_2) \in  R^\oplus }  \\
\end{array}$
\\[-.2cm]
\fine
\end{definition}

Note that, by definition, two markings are related by $R^\oplus$ only if they have the same size; 
in fact, the axiom states that
the empty marking is related to itself, while the rule, assuming by induction 
that $m_1$ and $m_2$ have the same size, ensures that $s_1 \oplus m_1$ and
$s_2 \oplus m_2$ have the same size.

\begin{proposition}\label{fin-k-add}
For each relation $R \subseteq S \times S$,  if $(m_1, m_2) \in R^\oplus$, 
then $|m_1| = |m_2|$.
\fine
\end{proposition}

Note also that the membership $(m_1, m_2) \in R^\oplus$ may be proved in several different 
ways, depending on the chosen order of the elements
of the two markings and on the definition of $R$. For instance, if $R = \{(s_1, s_3),$ 
$(s_1, s_4),$ $(s_2, s_3), (s_2, s_4)\}$,
then $(s_1 \oplus s_2, s_3 \oplus s_4) \in R^\oplus$ can be proved by means 
of the pairs $(s_1, s_3)$ and $(s_2, s_4)$,
as well as by means of $(s_1, s_4), (s_2, s_3)$.
An alternative way to define that two markings $m_1$ and $m_2$
are related by $R^\oplus$ is to state that $m_1$ can be represented as $s_1 \oplus s_2 \oplus \ldots \oplus s_k$, 
$m_2$ can be represented as $s_1' \oplus s_2' \oplus \ldots \oplus s_k'$ and $(s_i, s_i') \in R$ for $i = 1, \ldots, k$. 

\begin{proposition}\label{add-prop1}\cite{Gor17b,Gor25a}
For each relation $R \subseteq S \times S$, the following hold:
\begin{enumerate}
\item If $R$ is an equivalence relation, then $R^\oplus$ is an equivalence relation.
\item If $R_1 \subseteq R_2$, then $R_1^\oplus \subseteq R_2^\oplus$, i.e., the additive closure is monotone.
\item If $(m_1, m_2) \in R^\oplus$ and $(m_1', m_2') \in R^\oplus$,
then $(m_1 \oplus m_1', m_2 \oplus m_2') \in R^\oplus$, i.e., the additive closure is additive.
\item If $R$ is an equivalence relation, $(m_1 \oplus m_1', m_2 \oplus m_2') \in  R^\oplus$ and $(m_1, m_2) \in R^\oplus$,
then $(m_1', m_2') \in R^\oplus$, i.e., the additive closure is subtractive.
\item  If $(m_1, m_2) \in  R^\oplus$ and $m_1' \subseteq m_1$, then there exists $m_2' \subseteq m_2$
such that $(m_1', m_2') \in R^\oplus$ and $(m_1 \ominus m_1', m_2 \ominus m_2') \in R^\oplus$. \\[-1cm]
\end{enumerate}
\fine
\end{proposition}

We are now ready to introduce place bisimulation, which is a non-interleaving behavioral 
relation defined over the net places.

\begin{definition}\label{def-place-bis}{\bf (Place Bisimulation)}
Let $N = (S, A, T)$ be a P/T net. 
A {\em place bisimulation} is a relation
$R\subseteq S \times S$ such that if $(m_1, m_2) \in R^\oplus$
then
\begin{itemize}
\item $\forall t_1$ such that  $m_1[t_1\rangle m'_1$, $\exists t_2$ such that $m_2[t_2\rangle m'_2$ 
with $(\pre{t_1}, \pre{t_2}) \in R^\oplus$, $\ell(t_1) = \ell(t_2)$,  $(\post{t_1}, \post{t_2}) \in R^\oplus$ and, moreover, 
$(m_1', m_2') \in R^\oplus$; and symmetrically,
\item $\forall t_2$ such that  $m_2[t_2\rangle m'_2$, $\exists t_1$ such that $m_1[t_1\rangle m'_1$ 
with $(\pre{t_1}, \pre{t_2}) \in R^\oplus$, $\ell(t_1) = \ell(t_2)$,  $(\post{t_1}, \post{t_2}) \in R^\oplus$ and, moreover, 
$(m_1', m_2') \in R^\oplus$.
\end{itemize}
Two markings $m_1$ and $m_2$ are  {\em place bisimilar}, denoted by
$m_1 \sim_p m_2$, if there exists a place bisimulation $R$ such that $(m_1, m_2) \in R^\oplus$.
\fine
\end{definition}

Note that the two matching transitions $t_1$ and $t_2$ in the place bisimulation game must have essentially the same shape 
because $(i)$ $(\pre{t_1}, \pre{t_2}) \in R^\oplus$ implies that
$|\pre{t_1}| = |\pre{t_2}|$, then $(ii)$ $\ell(t_1) = \ell(t_2)$ and, finally, $(iii)$  
from $(\post{t_1}, \post{t_2}) \in R^\oplus$
we derive that $|\post{t_1}| = |\post{t_2}|$. 

\begin{proposition}\label{place-bis-eq}\index{Bisimulation!equivalence}\cite{ABS91,Gor21decid,Gor25a}
For each P/T net $N = (S, A, T)$, relation $\sim_p \; \subseteq  \mathcal{M}(S) \times  \mathcal{M}(S)$ is an
equivalence relation.
\fine
\end{proposition}

\begin{proposition}\label{place-int-prop}{\bf (Place bisimilarity is finer than interleaving bisimilarity)}
Given a P/T net $N = (S, A, T)$ and two markings $m_1, m_2$, if $m_1 \sim_p m_2$, then $m_1 \sim_{int} m_2$.
\proof
If $m_1 \sim_p m_2$, then there exists a place bisimulation $R$ such that $(m_1, m_2) \in R^\oplus$.
Note that if $R$ is a place bisimulation, then $R^\oplus$ is an interleaving bisimulation, and so $m_1 \sim_{int} m_2$.
\fine
\end{proposition}

\begin{figure}[t]
\centering
\begin{tikzpicture}[
every place/.style={draw,thick,inner sep=0pt,minimum size=6mm},
every transition/.style={draw,thick,inner sep=0pt,minimum size=4mm},
bend angle=45,
pre/.style={<-,shorten <=1pt,>=stealth,semithick},
post/.style={->,shorten >=1pt,>=stealth,semithick}
]
\def\eofigdist{3.2cm}
\def\eodist{0.55}

\node (a) [label=left:$a)\qquad \qquad $]{};

\node (q1) [place,tokens=1] [label={above:$s_1$} ] {};
\node (s1) [transition] [below left=\eodist of q1,label=left:$a\;$] {};
\node (s2) [transition] [below right=\eodist of q1,label=right:$\;b$] {};
\node (q2) [place] [below=\eodist of s1,label=left:$s_2\;$] {};
\node (q3) [place] [below=\eodist of s2,label=right:$\;s_3$] {};
\node (s3) [transition] [below=\eodist of q2, label=left:$b\;$] {};
\node (s4) [transition] [below=\eodist of q3,label=right:$\;a$] {};

\draw  [->, bend right] (q1) to (s1);
\draw  [->, bend left] (q1) to (s2);
\draw  [->] (s1) to (q2);
\draw  [->] (q2) to (s3);
\draw  [->] (s2) to (q3);
\draw  [->] (q3) to (s4);

\node (b) [right={3.1cm} of a,label=left:$b)\;\;$] {};

\node (p1) [place,tokens=1]  [right=\eofigdist of q1,label=above:$s_4$] {};
\node (p2) [place,tokens=1]  [right=\eodist of p1,label=above:$s_5$] {};
\node (t1) [transition] [below=\eodist of p1,label=left:$a\;$] {};
\node (t2)  [transition] [below=\eodist of p2,label=right:$\;b$] {};

\draw  [->] (p1) to (t1);
\draw  [->] (p2) to (t2);

\end{tikzpicture}
\caption{Two interleaving bisimilar, but non-place bisimilar, markings: $s_1$ and $s_4 \oplus s_5$}
\label{p>int-fig}
\end{figure}
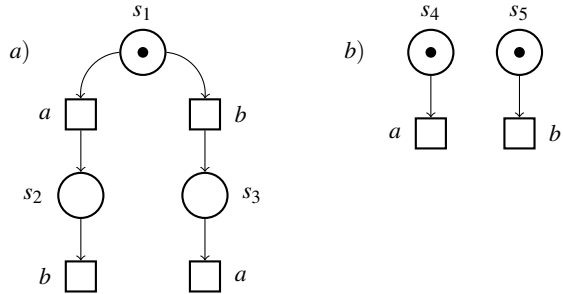

The implication above is strict, i.e., one can find two markings that are interleaving bisimilar 
but not place bisimilar. For instance, the markings $s_1$ (that is a singleton)
and $s_4 \oplus s_5$ (that has size 2) in Figure~\ref{p>int-fig} are not place bisimilar, as they have different size, but are clearly interleaving bisimilar.

\noindent
By Definition \ref{def-place-bis}, place bisimilarity $\sim_p$ can be defined as follows:

$\sim_p = \bigcup \{ R^\oplus \mid R \mbox{ is a place bisimulation}\}.$

\noindent
However, it is not true that 

$\sim_p = (\bigcup \{ R \mid R \mbox{ is a place bisimulation}\})^\oplus$

\noindent 
as the union of place bisimulations may not be a place bisimulation \cite{ABS91,Gor21decid,Gor25a}, 
so that its definition is not coinductive. 
Hence, the polynomial algorithms for checking bisimulation equivalence over LTSs (e.g., \cite{KS90,PT87})
cannot be adapted to compute $\sim_p$. Nonetheless, in \cite{Gor21decid,Gor25a} it is proved that $\sim_p$
can be decided in exponential time over the whole class of finite P/T Petri nets; a hint at this decidability proof is given in Section \ref{case-sec}.

%
\subsection{Structure-preserving Bisimulation}\label{sp-bis-sec}
%

In order to introduce this behavioral relation \cite{G15}, some auxiliary notation is needed.

\begin{definition}\label{link-def}{\bf (Link and linking)}\index{Link}\index{Linking}
Let $N = (S, A, T)$ be a P/T Petri net. 
A {\em link} is a pair $(s_1, s_2) \in S \times S$. 
A {\em linking} $l$
is a multiset of links, i.e.,
$l: S \times S \rightarrow \nat$ where $dom(l)$ is finite as $S$ is finite.
With abuse of notation, we denote by $\theta$ the empty linking, i.e., the empty multiset of links.
Given a set $L$ of links, we denote by $\mathcal{M}(L)$ the set of all the linkings over $L$. 
Given a linking $l$, the two projected markings $m_1$ and $m_2$ can be defined as

$m_1(s_1) = \pi_1(l)(s_1) =  \sum_{s_2 \in S} l(s_1, s_2)$ and 
$m_2(s_2) = \pi_2(l)(s_2) =  \sum_{s_1 \in S} l(s_1, s_2)$. 

\noindent
For instance, if $l = \{(s_1, s_2), (s_1', s_2')\}$, then $\pi_1(l) = s_1 \oplus s_1'$
and $\pi_2(l) = s_2 \oplus s_2'$. 
Note that the notation $(m_1, m_2) \in R^\oplus$ actually means that there exists a 
linking $l \in \mathcal{M}(R)$ such that
$\pi_1(l) = m_1$ and $\pi_2(l) = m_2$. As a matter of fact, a linking $l$ can be seen as a triple 
$(m_1, f, m_2)$, where $m_1$ and $m_2$ are
markings of the same size, and $f$ is a bijection between them.

We will use $l$, $\overline{l}$, $h$, $\overline{h}$, $c$ and $\overline{c}$, possibly indexed, to range over linkings.
As linkings are multisets, we can use the operations defined over multisets, such as union 
$\oplus$ and difference $\ominus$.
For instance, $(l \oplus c)(s_1, s_2) = l(s_1, s_2) + c(s_1, s_2)$. Moreover, a linking 
composed of a single link $(s_1, s_2)$
is denoted simply by its element, as we did for multisets in general. So, for instance, the 
linking $l = \{(s_1, s_2), (s_1', s_2')\}$
is often simply denoted by $l = (s_1, s_2) \oplus (s_1', s_2')$. 
\fine
\end{definition}

 \begin{definition}\label{def-sp1-bis}{\bf (Structure Preserving Bisimulation)}
Let $N = (S, A, T)$ be a P/T Petri net. 
A {\em structure-preserving bisimulation} (sp-bisimulation, for short)
is a set $R$ of linkings such that if $l \in R$, then $\forall c \subseteq l$, 
\begin{enumerate}
\item $\forall t_1$ such that $\pre{t_1} = \pi_1(c)$, there exist a transition $t_2$ 
such that $\ell(t_1) = \ell(t_2)$, $\pre{t_2} = \pi_2(c)$, and a linking $\overline{c}$ such that 
$\post{t_1} = \pi_1(\overline{c})$, $\post{t_2} = \pi_2(\overline{c})$ and $(l \ominus c) \oplus \overline{c} = \overline{l}  \in R$; 
\item $\forall t_2$ such that $\pre{t_2} = \pi_2(c)$, there exist a transition $t_1$ 
such that $\ell(t_1) = \ell(t_2)$, $\pre{t_1} = \pi_1(c)$, and a linking $\overline{c}$ such that 
$\post{t_1} = \pi_1(\overline{c})$, $\post{t_2} = \pi_2(\overline{c})$ and $(l \ominus c) \oplus \overline{c} = \overline{l} \in R$.
\end{enumerate}
Two markings $m_1$ and $m_2$ are  {\em structure-preserving} bisimulation equivalent (or, simply, sp-bisimilar), denoted by
$m_1 \sim_{sp} m_2$, if there exists a linking $l$ in a structure preserving bisimulation 
$R$ such that $m_1 = \pi_1(l)$
and $m_2 = \pi_2(l)$.
\fine
\end{definition}\index{Structure-preserving bisimulation!equivalence}
 
Note that the two matching transitions $t_1$ and $t_2$ in the structure-preserving bisimulation 
game must have essentially the same shape 
because $(i)$ $\pre{t_1} = \pi_1(c)$ and $\pre{t_2} = \pi_2(c)$ implies that
$|\pre{t_1}| = |\pre{t_2}|$, then $(ii)$ $\ell(t_1) = \ell(t_2)$ and, finally, $(iii)$  from $\post{t_1} = \pi_1(\overline{c})$ and $\post{t_2} = \pi_2(\overline{c})$
we derive that $|\post{t_1}| = |\post{t_2}|$. 

It is possible to prove that the definition of structure-preserving bisimulation is coinductive, as the 
union of sp-bisimulations is an sp-bisimulation. Moreover, for each P/T Petri net $N = (S, A, T)$, 
$\sim_{sp}$ is an equivalence relation. Moreover, $\sim_{sp}$ is decidable on bounded nets \cite{CG21a},
while the decidability on finite P/T Petri nets is open, as the observation in \cite{Esp98} does not apply to it.

Moreover, it is also possible to prove \cite{G15,Gor25a} that if $R$ is a place 
bisimulation, then $\mathcal{M}(R)$ is a
structure-preserving bisimulation, hence proving that $m_1 \sim_p m_2$ implies $m_1 \sim_{sp} m_2$. 
This implication is strict, as illustrated by the following example.

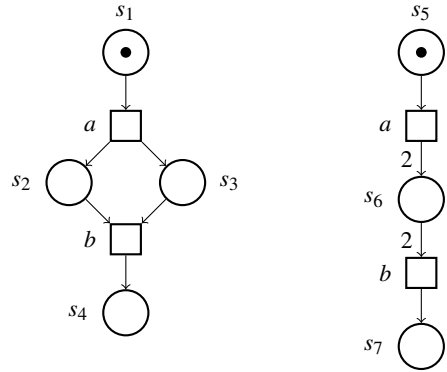
\begin{figure}[t]
\centering
\begin{tikzpicture}[
every place/.style={draw,thick,inner sep=0pt,minimum size=6mm},
every transition/.style={draw,thick,inner sep=0pt,minimum size=4mm},
bend angle=30,
pre/.style={<-,shorten <=1pt,>=stealth,semithick},
post/.style={->,shorten >=1pt,>=stealth,semithick}
]
\def\eofigdist{3.3cm}
\def\eodist{0.45cm}
\def\eodisty{1cm}

\node (p1) [place,tokens=1]  [label=above:$s_1$] {};
\node (t1) [transition] [below =\eodist of p1,label=left:$a\;$] {};
\node (p2) [place] [below left =\eodist of t1,label=left:$s_2\;$] {};
\node (p3) [place] [below right =\eodist of t1,label=right:$\;s_3$] {};
\node (t2) [transition] [below right=\eodist of p2,label=left:$b\;$] {};
\node (p4) [place] [below =\eodist of t2,label=left:$s_4\;$] {};

\draw  [->] (p1) to (t1);
\draw  [->] (t1) to (p2);
\draw  [->] (t1) to (p3);
\draw  [->] (p2) to (t2);
\draw  [->] (p3) to (t2);
\draw  [->] (t2) to (p4);

\node (p5) [place,tokens=1]  [right=\eofigdist of p1, label=above:$s_5$] {};
\node (t4) [transition] [below =\eodist of p5,label=left:$a\;$] {};
\node (p6) [place]  [below =\eodist of t4, label=left:$s_6\;$] {};
\node (t5) [transition] [below =\eodist of p6,label=left:$b\;$] {};
\node (p7) [place]  [below =\eodist of t5,label=left:$s_7\;$] {};

\draw  [->] (p5) to (t4);
\draw  [->] (t4) to node[auto,swap] {2} (p6);
\draw  [->] (p6) to node[auto,swap] {2} (t5);
\draw  [->] (t5) to (p7);

\end{tikzpicture}
\caption{Two sp-bisimilar nets, that are not place bisimilar}
\label{net-sp-place}
\end{figure}
     
\begin{example}\label{ex-sp1}
Consider the net in Figure \ref{net-sp-place} and the relation 

$R = \{(s_1, s_5), (s_2, s_6) \oplus (s_3, s_6),  (s_4, s_7)\}$

\noindent
composed of three linkings. It is not difficult to see that $R$ is a structure-preserving bisimulation.
However, no place bisimulation
relating $s_1$ and $s_5$  exists,
as this place relation would require the pair $(s_2, s_6)$, but $s_2$ and $s_6$ are not place bisimilar; 
in fact, $2 \cdot s_2$ is stuck, while $2 \cdot s_6$ can perform $b$.
\fine
\end{example}

%
\section{Causality-based Semantics}\label{causal-petri-sec}
%

We outline some definitions, adapted from the literature
(cf., e.g., \cite{GR83,BD87,Old,G15,Gor22}).

\begin{definition}\label{acyc-def}{\bf (Acyclic net)}
A P/T net $N = (S, A, T)$ is
 {\em acyclic} if its flow relation $F$ is acyclic (i.e., $\not \exists x$ such that $x F^+ x$, 
 where $F^+$ is the transitive closure of $F$).
 \fine
\end{definition}

The causal semantics of a marked P/T net is defined by a class of particular acyclic safe nets, 
where places are not branched (that is, they have at most one incoming and one outgoing edge, hence the nets represent a single run) and all arcs have weight 1. 
This kind of net is called {\em causal net} (or {\em occurrence net}) \cite{GR83,BD87,Old,G15}. 
We use the name $C$ (possibly indexed) to denote a causal net, the set $B$ to denote its 
places (called {\em conditions}), the set $E$ to denote its transitions 
(called {\em events}), and
$L$ to denote its labels.

\begin{definition}\label{causalnet-def}{\bf (Causal net)}
A causal net is a finite marked P/T net $C(\mathsf{m}_0) = (B,L, 
E,  \mathsf{m}_0)$ satisfying
the following conditions:
\begin{enumerate}
\item $C$ is acyclic;
\item $\forall b \in B \; \; | \pre{b} | \leq 1\, \wedge \, | \post{b} | \leq 1$ (i.e., the places are not branched);
\item  $ \forall b \in B \; \; \mathsf{m}_0(b)   =  \begin{cases}
 1 & \mbox{if $\; \pre{b} = \emptyset$}\\ 
  0  & \mbox{otherwise;}   
   \end{cases}$\\
\item $\forall e \in E \; \; \pre{e}(b) \leq 1 \, \wedge \, \post{e}(b) \leq 1$ for all $b \in B$ (i.e., all the arcs have weight $1$).
\end{enumerate}
We denote by $Min(C)$ the set $\mathsf{m}_0$, and by $Max(C)$ the set
$\{b \in B \mid \post{b} = \emptyset\}$.
\fine
\end{definition}

Note that any reachable marking of a causal net is a set, i.e., 
this net is {\em safe}; in fact, the initial marking is a set and, 
assuming by induction that a reachable marking $\mathsf{m}$ is a set and enables $e$, i.e., 
$\mathsf{m}[e\rangle \mathsf{m}'$,
then also
$\mathsf{m}' =  (\mathsf{m} \ominus \pre{e}) \oplus \post{e}$ is a set, 
as the net is acyclic and because
of the condition on the shape of the post-set of $e$ (weights can only be $1$).

 As the initial marking of a causal net is fixed by its shape (according to item $3$ of 
Definition \ref{causalnet-def}), in the following, in order to make the 
 notation lighter, we often omit the indication of the initial marking, 
 so that the causal 
 net $C(\mathsf{m}_0)$ is denoted by $C$.

\begin{definition}\label{trans-causal}{\bf (Moves of a causal net)}
Given two causal nets $C = (B, L, E,  \mathsf{m}_0)$
and $C' = (B', L, E',  \mathsf{m}_0)$, we say that $C$
moves in one step to $C'$ through $e$, denoted by
$C [e\rangle C'$, if $\; \pre{e} \subseteq Max(C)$, $E' = E \cup \{e\}$
and $B' = B \cup \post{e}$. 
\fine
\end{definition}

\noindent 
Note that a move $C [e\rangle C'$ extends the causal net $C$ to $C'$ by adding the maximal event $e$.

\begin{definition}\label{folding-def}{\bf (Folding and process)}
A {\em folding} from a causal net $C = (B, L, E, \mathsf{m}_0)$ into a P/T net system
$N(m_0) = (S, A, T, m_0)$ is a function $\rho: B \cup E \to S \cup T$, which is type-preserving, i.e., 
such that $\rho(B) \subseteq S$ and $\rho(E) \subseteq T$, satisfying the following:
\begin{itemize}
\item $L = A$ and $\mathsf{l}(e) = \ell(\rho(e))$ for all $e \in E$ (i.e., foldings preserve labels of events);
\item $\rho(\mathsf{m}_0) = m_0$, i.e., $m_0(s) = | \rho^{-1}(s) \cap \mathsf{m}_0 |$ (foldings preserve the initial marking);
\item $\forall e \in E, \rho(\pre{e}) = \pre{\rho(e)}$, i.e., $\rho(\pre{e})(s) = | \rho^{-1}(s) \cap \pre{e} |$
for all $s \in S$ (foldings and pre-set commute);
\item $\forall e \in E, \, \rho(\post{e}) = \post{\rho(e)}$,  i.e., $\rho(\post{e})(s) = | \rho^{-1}(s) \cap \post{e} |$
for all $s \in S$ (foldings and post-set commute).
\end{itemize}
A pair $(C, \rho)$, where $C$ is a causal net and $\rho$ a folding from  
$C$ to a net system $N(m_0)$, is a {\em process} of $N(m_0)$. 
\fine
\end{definition}

\begin{definition}\label{trans-process}{\bf (Moves of a process)}
Let $N(m_0) = (S, A, T, m_0)$ be a P/T net system 
and let $(C_i, \rho_i)$, for $i = 1, 2$, be two processes of $N(m_0)$.
We say that $(C_1, \rho_1)$
moves in one step to $(C_2, \rho_2)$ through $e$, denoted by
$(C_1, \rho_1) \deriv{e} (C_2, \rho_2)$, if $C_1 [e\rangle C_2$
and $\rho_1 \subseteq \rho_2$.
\noindent
If $\pi_1 = (C_1, \rho_1)$ and $\pi_2 = (C_2, \rho_2)$, we denote
the move as $\pi_1 \deriv{e} \pi_2$.
\fine
\end{definition}

\begin{definition}\label{po-process-def}{\bf (Partial orders of events)}
From a causal net $C = (B, L, E, \mathsf{m}_0)$,  we can 
extract the {\em partial order of its events}
$\mathsf{E}_{\mathsf{C}} = (E, \preceq)$,
where $e_1 \preceq e_2$ if there is a path in the net from $e_1$ to $e_2$, i.e., if $e_1 \mathsf{F}^* e_2$, where
$\mathsf{F}^*$ is the reflexive and transitive closure of
$\mathsf{F}$, which is the flow relation for $C$.

Two partial orders $(E_1, \preceq_1)$ and $(E_2, \preceq_2)$ are isomorphic 
if there is a label-preserving, order-preserving  bijection $g: E_1 \to E_2$, i.e., such that
$\mathsf{l}_1(e) = \mathsf{l}_2(g(e))$ and $e \preceq_1 e'$ if and only if $g(e) \preceq_2 g(e')$.
We also say that a map $g$ is an
{\em event isomorphism} between the causal nets 
$C_1$ and 
$C_2$ if it is an isomorphism between their associated 
partial orders $\mathsf{E}_{C_1}$
and $\mathsf{E}_{C_2}$.
\fine
\end{definition}

%
\subsection{Causal-net Bisimulation}\label{cn-bis-sec}
%

\begin{definition}\label{cn-bis-def}{\bf (Causal-net bisimulation)}\cite{G15,Gor22}
Let $N = (S, A, T)$ be a P/T net. A {\em causal-net bisimulation} 
is a relation $R$, composed of 
triples of the form $(\rho_1, C, \rho_2)$, where, for $i = 1, 2$, $(C, \rho_i)$ is a process 
of $N(m_{0_i})$ for some $m_{0_i}$,
such that if $(\rho_1, C, \rho_2) \in R$ then

\begin{itemize}
\item[$i)$] 
$\forall t_1, C', \rho_1'$ s.t.  $(C, \rho_1) \deriv{e} (C', \rho_1')$,
where $\rho_1'(e) = t_1$,
$\exists t_2, \rho_2'$ s.t.
$(C, \rho_2) \deriv{e} (C', \rho_2')$,
where $\rho_2'(e) = t_2$, and
$(\rho'_1, C', \rho'_2) \in R$; and symmetrically,

\item[$ii)$] $\forall t_2, C', \rho_2'$ s.t. $(C, \rho_2) \deriv{e} (C', \rho_2')$,
where $\rho_2'(e) = t_2$,
$\exists t_1, \rho_1'$ s.t.
$(C, \rho_1) \deriv{e} (C', \rho_1')$,
where $\rho_1'(e) = t_1$, and
$(\rho'_1, C', \rho'_2) \in R$.
\end{itemize}
Two markings $m_{1}$ and $m_2$ of $N$ are cn-bisimilar (or causal-net bisimulation equivalent), 
denoted by $m_{1} \sim_{cn} m_{2}$, 
if there exists a causal-net bisimulation $R$ containing a triple $(\rho^0_1, C^0, \rho^0_2)$, 
where $C^0$ contains no events (i.e., $E^0 = \emptyset$) and 
$\rho^0_i(Min( C^0))  = $ $\rho^0_i(Max( C^0)) = m_i\;$ for $i = 1, 2$.
\fine
\end{definition}

\begin{theorem}\label{sp=cn-th}{\bf (Causal-net bisimilarity and sp-bisimilarity coincide)}\cite{G15,Gor25a}
Let $N = (S, A, T)$ be a P/T net. We have that $m_1 \sim_{sp} m_2$ if and only if 
$m_1 \sim_{cn} m_2$.
\fine
\end{theorem}

Since $\sim_{sp}$ and $\sim_{cn}$ coincide, also causal-net bisimilarity is decidable 
on bounded nets \cite{CG21a}.

\begin{definition}\label{hcn-bis-def}{\bf (Hereditary causal-net bisimulation)}
Let $N = (S, A, T)$ be a P/T net. A {\em hereditary} causal-net bisimulation 
is a causal net bisimulation relation $R$
such that if $(\rho_1, C, \rho_2) \in R$, then 
\begin{itemize}
\item[$i)$] 
$\forall t_1, C', \rho_1'$ s.t.  $(C', \rho_1') \deriv{e} (C, \rho_1)$,
where $\rho_1(e) = t_1$,
$\exists t_2, \rho_2'$ s.t.
$(C', \rho_2') \deriv{e} (C, \rho_2)$,
where $\rho_2(e) = t_2$, and
$(\rho'_1, C', \rho'_2) \in R$; and symmetrically,

\item[$ii)$] 
$\forall t_2, C', \rho_2'$ s.t. $(C', \rho_2') \deriv{e} (C, \rho_2)$,
where $\rho_2(e) = t_2$,
$\exists t_1, \rho_1'$ s.t.
$(C', \rho_1') \deriv{e} (C, \rho_1)$,
where $\rho_1(e) = t_1$, and
$(\rho'_1, C', \rho'_2) \in R$.
\end{itemize}
Two markings $m_{1}$ and $m_2$ of $N$ are hcn-bisimilar (or hereditary causal-net bisimulation equivalent), 
denoted by $m_{1} \sim_{hcn} m_{2}$, 
if there exists a hereditary causal-net bisimulation $R$ containing a triple $(\rho^0_1, C^0, \rho^0_2)$, 
where $C^0$ contains no events (i.e., $E^0 = \emptyset$) and 
$\rho^0_i(Min( C^0))  = \rho^0_i(Max( C^0)) = m_i\;$ for $i = 1, 2$.
\fine
\end{definition}

Hcn-bisimulation requires to match transitions \emph{moving to} equivalent processes. 
This corresponds to match the reverse of such transitions, which are, as usual in bisimulations, \emph{from} equivalent processes.

\begin{theorem}\label{cn=hcn-th}{\bf (Causal-net bisimilarity is self-reversible)}
Let $N = (S, A, T)$ be a P/T net. We have that $m_1 \sim_{cn} m_2$ if and only if 
$m_1 \sim_{hcn} m_2$.
\proof
By definition, a hereditary causal-net bisimulation is a causal-net bisimulation,
hence we only have to prove
the implication from left to right.

Assume $R$ is a causal net bisimulation justifying that $m_1 \sim_{cn} m_2$; this means that
$R$ contains a triple $(\rho^0_1, C^0, \rho^0_2)$, 
where $C^0$ contains no events (i.e., $E^0 = \emptyset$) and 
$\rho^0_i(Min( C^0)) $ $ = \rho^0_i(Max( C^0)) = m_i\;$ for $i = 1, 2$.
We want to prove, by induction on the length $n \geq 1$ of the computations originating from configuration $(C^0, \rho^0_1)$,
that the hcn-conditions are satisfied. The case when $(C^0, \rho^0_2)$ moves first is symmetric,
and so omitted.

The base case is $n = 1$. Hence, assume $(C^0, \rho^0_1) \deriv{e_1} (C^1, \rho^1_1)$. 
As $(\rho^0_1, C^0, \rho^0_2) \in R$, there exists $(C^0, \rho^0_2) \deriv{e_1} (C^1, \rho^1_2)$
such that $(\rho^1_1, C^1, \rho^1_2) \in R$. If we backtrack event $e_1$, we end up to $(C^0, \rho^0_1)$
and $(C^0, \rho^0_2)$, respectively, such that $(\rho^0_1, C^0, \rho^0_2) \in R$. So, this case is trivial.
Now assume that 

$(C^0, \rho^0_1) \deriv{e_1} (C^1, \rho^1_1) \deriv{e_2} \ldots  \deriv{e_{n-2}}
(C^{n-2}, \rho^{n-2}_1) \deriv{e_{n-1}} (C^{n-1}, \rho^{n-1}_1)\deriv{e_{n}} (C^n, \rho^n_1)$,

\noindent
with $n \geq 2$. By induction, we know that there exists a computation 

$(C^0, \rho^0_2) \deriv{e_1} (C^1, \rho^1_2) \deriv{e_2} \ldots \deriv{e_{n-2}} (C^{n-2}, \rho^{n-2}_2) \deriv{e_{n-1}} (C^{n-1}, \rho^{n-1}_2)$,

\noindent
such that $(\rho^i_1, C^i, \rho^i_2) \in R$ for $i = 1, \ldots, n-1$ and, moreover, that is reversible, i.e., it satisfies the 
hcn-conditions by backtracking events.

As $(\rho^{n-1}_1, C^{n-1}, \rho^{n-1}_2) \in R$, a transition
$(C^{n-1}, \rho^{n-1}_2) \deriv{e_n} (C^n, \rho^n_2)$ exists
such that $(\rho^n_1, C^n, \rho^n_2) \in R$. 
If $e_{n}$ causally depends on $e_{n-1}$, then
we can only backtrack $e_n$, going back to the pairs $(C^{n-1}, \rho^{n-1}_1)$ and $(C^{n-1}, \rho^{n-1}_2)$, with $(\rho^{n-1}_1, C^0, \rho^{n-1}_2) \in R$, so that this case is trivial, too.
On the contrary, if $e_n$ and $e_{n-1}$ are causally independent, then we could also backtrack 
$e_{n-1}$.
Therefore, we get:

$(C^{n-2}, \rho^{n-2}_1) \deriv{e_{n}} (\underline{C}^{n-1}, \underline{\rho}^{n-1}_1) \deriv{e_{n-1}} (C^n, \rho^n_1)$ and

$(C^{n-2}, \rho^{n-2}_2) \deriv{e_{n}} (\underline{C}^{n-1}, \underline{\rho}^{n-1}_2) \deriv{e_{n-1}} (C^n, \rho^n_2)$.

Therefore, it remains to argue that $(\underline{\rho}^{n-1}_1, \underline{C}^{n-1}, \underline{\rho}^{n-1}_2) \in R$.
However, note that, given $(\rho^n_1, C^n, \rho^n_2) \in R$  and $(\rho^{n-2}_1, C^{n-2}, \rho^{n-2}_2) \in R$, this is due to the 
fact that the foldings on these independent events are already fixed in the target configurations
$(C^n, \rho^n_1)$ and $(C^n, \rho^n_2)$, 
so that they are also fixed accordingly in the intermediate configurations $(\underline{C}^{n-1}, \underline{\rho}^{n-1}_1)$
and $(\underline{C}^{n-1}, \underline{\rho}^{n-1}_2)$. 
Finally, induction can be invoked to conclude that also the computations
of length $n-1$ ending with $(\underline{C}^{n-1}, \underline{\rho}^{n-1}_1)$
and $(\underline{C}^{n-1}, \underline{\rho}^{n-1}_2)$, respectively, can match their corresponding backward transitions.
\fine
\end{theorem}

Since place bisimilarity $\sim_p$ implies structure-preserving bisimilarity $\sim_{sp}$, by Theorem \ref{sp=cn-th} we get the
expected result that place bisimilarity $\sim_p$ implies causal-net bisimilarity $\sim_{cn}$. (The reverse 
implication does not hold, as
illustrated in Figure \ref{net-sp-place}.)
Below we provide a direct proof of this result
as this proof is instrumental to the following example, that clarifies the differences between place bisimulation and causal-net bisimulation.

\begin{theorem}\label{place-bis>cn-bis}
For each P/T net $N = (S, A, T)$, if $m_1 \sim_p m_2$, then $m_1 \sim_{cn} m_2$.
\proof
If $m_1 \sim_{p} m_2$, then  a place bisimulation $R_1$ exists with
$(m_1, m_2) \in R_1^\oplus$.
Consider 
    \begin{equation*} \label{R2}
        \begin{split}
        R_2 \overset{def}{=} \lbrace (\rho_1, C, \rho_2) | & (C, \rho_1) \text{ is a process of $N(m_{1})$ and } 
        (C, \rho_2) \text{ is a process of $N(m_{2})$ and} \\
        & \forall b \in Max(C) \; (\rho_1(b), \rho_2(b)) \in R_1
         \rbrace .
        \end{split}
    \end{equation*}

\noindent
We want to prove that $R_2$ is a causal-net bisimulation.
First of all, consider a triple of the form $(\rho_1^0, C^0, \rho_2^0)$,
where $C^0$ is the causal net without events and $\rho_1^0, \rho_2^0$ are chosen in such a way that, 
not only  $\rho_i^0(Min(C^0)) = \rho_i^0(Max(C^0)) = m_i$ for $i= 1, 2$, but also that 
$(\rho_1^0(b), \rho_2^0(b)) \in R_1$ for all $b \in Max(C^0)$, which is really possible because we know that $(m_1, m_2) \in R_1^\oplus$.
Then $(\rho_1^0, C^0, \rho_2^0)$ must belong to $R_2$,
because $(C^0, \rho_i^0)$ is a process of $N(m_i)$, for $i=1, 2$ and, 
by construction, $(\rho_1^0(b), \rho_2^0(b)) \in R_1$ for all $b \in Max(C^0)$. 
Hence, if $R_2$ is a causal-net bisimulation, then the triple 
$(\rho_1^0, C^0, \rho_2^0) \in R_2$ ensures that $m_1 \sim_{cn} m_2$. 
Now, assume $(\rho_1, C, \rho_2) \in R_2$. 
In order for $R_2$ to be a causal-net bisimulation,
we must prove that
\begin{enumerate}
\item
$\forall t_1, C', \rho_1'$ s.t. $(C, \rho_1) \deriv{e} (C', \rho_1')$,
where $\rho_1'(e) = t_1$,
$\exists t_2, \rho_2'$ s.t.
$(C, \rho_2) \deriv{e} (C', \rho_2')$,
where $\rho_2'(e) = t_2$, and
$(\rho'_1, C', \rho'_2) \in R_2$;
\item symmetrically, if $(C, \rho_2)$ moves first.
\end{enumerate}

Assume $(C, \rho_1) \deriv{e} (C', \rho_1')$ with $\rho_1'(e) = t_1$.
Recall that if $(\rho_1, C, \rho_2)  \in R_2$ then  for all $b \in Max(C)$ we have $(\rho_1(b),$ $ \rho_2(b)) \in R_1$.
This means that $(\rho_1(\pre{e}), \rho_2(\pre{e})) \in R_1^\oplus$. As $R_1$ is a place bisimulation,
to the move $\rho_1(\pre{e})[t_1\rangle \post{t_1} $, marking  $\rho_2(\pre{e})$ can respond
with $\rho_2(\pre{e})[t_2\rangle \post{t_2}$ for some suitable transition $t_2$
such that $(\pre{t_1}, \pre{t_2}) \in R_1^\oplus$, $\ell(t_1) = \ell(t_2)$ and $(\post{t_1}, \post{t_2}) \in R_1^\oplus$.
Therefore, since $t_1$ and $t_2$ have the same shape,
it is possible to derive $(C, \rho_2) \deriv{e} (C', \rho_2')$,
where $\rho_2'$ is chosen in such a way that, not only $\rho_2'(e) = t_2$, but also that $(\rho_1'(b), \rho_2'(b)) \in R_1$ for each $b \in \post{e}$ 
(which is really possible because $(\post{t_1}, \post{t_2}) \in R_1^\oplus$).
Since for all $b \in (Max(C) \ominus \pre{e})$ we already know that  $(\rho_1(b),$ $ \rho_2(b)) \in R_1$,
it follows that
for all $b' \in Max(C')$ it holds that $(\rho_1'(b'), \rho_2'(b')) \in R_1$.
As a consequence, we also have that $(\rho'_1, C', \rho'_2) \in R_2$.

The case when $(C, \rho_2)$ moves first is symmetrical and therefore omitted.
Thus, $R_2$ is a causal-net bisimulation and, since $(\rho_1^0, C^0, \rho_2^0) \in R_2$, we have
$m_1 \sim_{cn} m_2$.
\fine
\end{theorem}
 
We have to be a bit careful about the result we have just proved. It states that for each computation originating from 
marking $m_1$, we can {\em choose}, among the many computations originating from $m_2$ that do match 
(according to the place bisimulation definition), 
the one that generates the same causal net. The following example clarifies this issue.
 
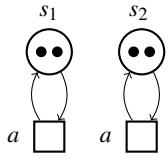
\begin{figure}[!t]
\centering
\begin{tikzpicture}[
every place/.style={draw,thick,inner sep=0pt,minimum size=6mm},
every transition/.style={draw,thick,inner sep=0pt,minimum size=4mm},
bend angle=30,
pre/.style={<-,shorten <=1pt,>=stealth,semithick},
post/.style={->,shorten >=1pt,>=stealth,semithick}
]
\def\eofigdist{3.3cm}
\def\eodist{0.6cm}
\def\eodisty{1cm}

\node (p1) [place,tokens=2]  [label=above:$s_1$] {};
\node (t1) [transition] [below =\eodist of p1,label=left:$a\;$] {};
\node (p2) [place,tokens=2] [right =\eodist of p1,label=above:$s_2\;$] {};
\node (t2) [transition] [below =\eodist of p2,label=left:$a\;$] {};

\draw   [->, bend left] (p1) to (t1);
\draw   [->, bend left] (t1) to (p1);
\draw   [->, bend left] (p2) to (t2);
\draw   [->, bend left] (t2) to (p2);

\end{tikzpicture}
\caption{A simple net with replicas}
\label{net-no-good-place}
\end{figure}

\begin{example}\label{no-good-ex}
Consider the net in Figure \ref{net-no-good-place}, where we call by $t_i$ the transition $(s_i, a, s_i)$ for $i = 1, 2$.
Consider also the relation $R = \{(s_1, s_2)\}$, that is clearly a place bisimulation,
and the two markings $2 \cdot s_1$ and $2 \cdot s_2$. Of course, $(2 \cdot s_1, 2 \cdot s_2) \in R^\oplus$
because $(s_1, s_2) \in R$, and so, by 
Theorem \ref{place-bis>cn-bis}, $2 \cdot s_1$ and $2 \cdot s_2$ are causal-net bisimilar markings.
However, it may happen that
computations, that do match (according to the place bisimulation definition), originate different causal nets.

For instance, if $2 \cdot s_1 [t_1 \rangle 2 \cdot s_1 [t_1 \rangle 2 \cdot s_1$ and the two occurrences 
of $t_1$ are causally related (because the latter consumes the token produced by the former),
then, according to the place bisimulation definition, marking $2 \cdot s_2$ can 
also reply
with the 
sequence $t_2 t_2$,
where the two occurrences of $t_2$ are instead independent (as each consumes a different token). 
Similarly, if the two occurrences of $t_1$ are independent, then $2 \cdot s_2$ 
can reply also with the sequence $t_2 t_2$,
where the two occurrences of $t_2$ are causally related. 
Actually, whether the two transitions are causally related or independent is immaterial for the place bisimulation, but visible in the causal net bisimulation.

Nonetheless, this observation does not invalidate the result of Theorem \ref{place-bis>cn-bis}, because to the computation 
$2 \cdot s_1 [t_1 \rangle 2 \cdot s_1 [t_1 \rangle 2 \cdot s_1$ where the two occurrences of $t_1$ are causally related,
$2 \cdot s_2$ can reply with $2 \cdot s_2 [t_2 \rangle 2 \cdot s_2 [t_2\rangle 2 \cdot s_2$ where the two occurrences of $t_2$ are causally dependent; 
while when the two occurrences of $t_1$ are independent,
then $2 \cdot s_2$ can reply with the transition sequence $t_2t_2$ of independent occurrences of $t_2$,
so that the matched computations generate the same causal net. In other words, if the second transition 
$2 \cdot s_1 [t_1 \rangle 2 \cdot s_1$ may 
represent both a causally dependent occurrence or and independent occurrence of $t_1$, then the matching $2 \cdot s_2 [t_2 \rangle 2 \cdot s_2$, according to the place bisimulation game, can both represent a causally dependent occurrence or and independent 
occurrence of $t_2$, too.

In fact, the crucial fact ensuring the correctness
of Theorem \ref{place-bis>cn-bis} is that for each computation originating from 
marking $2 \cdot s_1$, we can {\em choose}, among the many computations from $2 \cdot s_2$
that do match (according to the place bisimulation definition), 
the one, that for sure exists, generating the same causal net.
Similar considerations hold also for the left-to-right implication in Theorem \ref{sp=cn-th}.
\fine
\end{example}

%
\subsection{Fully Concurrent Bisimulation}\label{fc-bis-sec}
%

Causal-net bisimilarity observes not only the partial order of generated events, but also the size of the
current distributed state, as it observes the causal nets. A coarser behavioral equivalence, ignoring the size 
of the distributed state and the shape of transitions, is
{\em fully concurrent bisimulation} (fc-bisimulation, for short) \cite{BDKP91}, whose definition
was inspired by previous equivalence notions on other concurrency models:
{\em history-preserving bisimulation}, originally defined in \cite{RT88} under the name of {\em behavior-structure bisimulation}, and 
then elaborated on in \cite{vGG89} (who called it by this name) and also independently defined in \cite{DDM89} 
(who called it {\em mixed ordering bisimulation}). Its definition follows.

\begin{definition}\label{sfc-bis-def}{\bf (Fully concurrent bisimulation)}
Given a P/T net $N = (S, A, T)$, a {\em fully concurrent bisimulation} (fc-bisimulation, for short)
is a relation $R$, composed of 
triples of the form $(\pi_1, g, \pi_2) $, where, for $i = 1,2$, 
$\pi_i = (C_i, \rho_i)$ is a process of $N(m_{0i})$ for some $m_{0i}$ and
$g$ is an event isomorphism between $\mathsf{E}_{C_1}$ and $\mathsf{E}_{C_2}$, such that  
if $(\pi_1, g, \pi_2) \in R$ then
\begin{enumerate}
\item 
$\forall t_1, \pi_1'$ such that $\pi_1 \deriv{e_1} \pi_1'$ with $\rho_1'(e_1) = t_1$, 
$\exists t_2, \pi_2', g'$ such that
\begin{enumerate} 
\item $\pi_2 \deriv{e_2} \pi_2'$ with $\rho_2'(e_2) = t_2$;
\item $g' = g \cup \{(e_1, e_2)\}$, and finally,
\item $(\pi_1', g', \pi_2') \in R$;
\end{enumerate}
\item and symmetrically, if $\pi_2$ moves forward first.
\end{enumerate}
Two markings $m_1, m_2$ are fully concurrent bisimilar,
denoted $m_1 \sim_{fc} m_2$, if a fully concurrent bisimulation R exists,
containing a triple $(\pi^0_1, \emptyset, \pi^0_2)$ where 
$\pi^0_i = (C^0_i, \rho^0_i)$ is such that 
$C^0_i$ contains no events (i.e., $E^0_i = \emptyset$) and 
$\rho^0_i(Min(C^0_i))  = \rho^0_i(Max(C^0_i))$ $ = m_i\;$ for $i = 1, 2$.
\fine
\end{definition}

\begin{definition}\label{hsfc-bis-def}{\bf (Hereditary fully concurrent bisimulation)}
Given a P/T net $N = (S, A, T)$, a {\em hereditary} fully concurrent bisimulation
(hfc-bisimulation, for short) is an fc-bisimulation such that  
if $(\pi_1, g, \pi_2) \in R$, then 
\begin{enumerate}
\item
$\forall t_1, \pi_1'$ such that $\pi_1' \deriv{e_1} \pi_1$ with $\rho_1(e_1) = t_1$, 
$\exists t_2, \pi_2', g'$ such that
$(i)$ $\pi_2' \deriv{e_2} \pi_2$ with $\rho_2(e_2) = t_2$;
 $(ii)$ $g' = g \setminus \{(e_1, e_2)\}$, and finally,
$(iii)$ $(\pi_1', g', \pi_2') \in R$;
\item and symmetrically, if $\pi_2$ moves backward first.
\end{enumerate}
Two markings $m_1, m_2$ are hereditary fully concurrent bisimilar,
denoted $m_1 \sim_{hfc} m_2$, if a hereditary fully concurrent bisimulation R exists,
containing a triple $(\pi^0_1, \emptyset, \pi^0_2)$ where 
$\pi^0_i = (C^0_i, \rho^0_i)$ is such that 
$C^0_i$ contains no events (i.e., $E^0_i = \emptyset$) and 
$\rho^0_i(Min(C^0_i))  = \rho^0_i(Max(C^0_i))$ $ = m_i\;$ for $i = 1, 2$.
\fine
\end{definition}

Of course, $\sim_{cn} = \sim_{hcn}$ is finer than $\sim_{hfc}$. 
This can be proved by observing that 
if $R_1$ is a causal-net bisimulation,
then 
$R_2 = \{(C, \rho_1), id, (C, \rho_2) $ $\mid (\rho_1, C, \rho_2) \in R_1\}$, 
 where $id$ is 
the identity function on $E$, is an hfc-bisimulation.

We can summarize the relationships among the behavioral equivalences discussed in this paper as follows: 
$\sim_p \subset \sim_{hcn} = \sim_{cn} = \sim_{sp} \subset \sim_{hfc} \subset \sim_{fc} \subset \sim_{int}$. In words,
place bisimilarity is the finest behavioral equivalence, which
is finer than (hereditary) causal-net bisimilarity (or, equivalently, 
sp-bisimilarity), in turn finer than hereditary fully concurrent bisimilarity,
in turn finer than fully concurrent bisimilarity, in turn finer than interleaving bisimilarity. All inclusions are strict.
An example showing that $\sim_p$ is finer than $\sim_{cn}$ is outlined in Figure \ref{net-sp-place};
examples showing that causal-net bisimilarity is finer than hereditary fully concurrent bisimilarity are 
outlined in Figure \ref{net-fc-place} and Figure \ref{cn-vs-icn-fig}.

\begin{figure}[t]
\centering

\begin{tikzpicture}[
every place/.style={draw,thick,inner sep=0pt,minimum size=6mm},
every transition/.style={draw,thick,inner sep=0pt,minimum size=4mm},
bend angle=42,
pre/.style={<-,shorten <=1pt,>=stealth,semithick},
post/.style={->,shorten >=1pt,>=stealth,semithick}
]
\def\eofigdist{6.5cm}
\def\eodist{0.4cm}
\def\eodistz{1cm}
\def\eodisty{1.6cm}
\def\eodistw{3cm}

\node (p1) [place,tokens=1]  [label=above:$s_1$] {};
\node (t1) [transition] [left=\eodistz of p1, label=above:$a_1$]{};
\node (t2) [transition] [right=\eodistz of p1, label=above:$b_2$]{};

\node (p2) [place]  [below=\eodist of t1,label=left:$s_2$] {};
\node (t3) [transition] [below=\eodist of p2,label=left:$c$] {};

\node (p3) [place]  [below=\eodist of t2,label=right:$s_3$] {};
\node (t4) [transition] [below=\eodist of p3,label=right:$d$] {};

\node (p4) [place,tokens=1] [below left=\eodisty of t1,label=right:$s_4$] {};
\node (p5) [place,tokens=1] [below right=\eodisty of t2,label=left:$s_5$] {};
\node (p6) [place,tokens=1] [below =\eodistw of p1,label=below:$s_6$] {};

\node (t5) [transition] [left=\eodistz of p6,label=below:$a_2$] {};
\node (t6) [transition] [right=\eodistz of p6,label=below:$b_1$] {};

\draw  [->] (p1) to (t1);
\draw  [->] (p1) to (t2);
\draw  [->] (t1) to (p2);
\draw  [->] (t2) to (p3);
\draw  [->] (p2) to (t3);
\draw  [->] (p3) to (t4);
\draw  [->, bend left] (p4) to (t1);
\draw  [->, bend right] (p4) to (t4);
\draw  [->, bend right] (p5) to (t2);
\draw  [->, bend left] (p5) to (t3);
\draw  [->] (p6) to (t5);
\draw  [->] (p6) to (t6);
\draw  [->, bend right] (p4) to (t5);
\draw  [->, bend left] (p5) to (t6);

\node (q1) [place,tokens=1]  [right=\eofigdist of p1,label=above:$r_1$] {};
\node (v1) [transition] [left=\eodistz of q1, label=above:$a_1'$]{};
\node (v2) [transition] [right=\eodistz of q1, label=above:$b_2'$]{};

\node (q2) [place]  [below=\eodist of v1,label=left:$r_2$] {};
\node (v3) [transition] [below=\eodist of q2,label=left:$c$] {};

\node (q4) [place,tokens=1] [below left=\eodisty of v1,label=right:$r_4$] {};
\node (q5) [place,tokens=1] [below right=\eodisty of v2,label=left:$r_5$] {};
\node (q6) [place,tokens=1] [below =\eodistw of q1,label=below:$r_6$] {};

\node (v5) [transition] [left=\eodistz of q6,label=below:$a_2'$] {};
\node (v6) [transition] [right=\eodistz of q6,label=below:$b_1'$] {};

\node (q3) [place]  [above=\eodist of v6,label=left:$r_3$] {};
\node (v4) [transition] [above=\eodist of q3,label=left:$d$] {};

\draw  [->] (q1) to (v1);
\draw  [->] (q1) to (v2);
\draw  [->] (q2) to (v3);
\draw  [->] (v1) to (q2);
\draw  [->] (v6) to (q3);
\draw  [->] (q3) to (v4);
\draw  [->, bend left] (q4) to (v1);
\draw  [->, bend right] (q4) to (v4);
\draw  [->, bend right] (q5) to (v2);
\draw  [->, bend right] (q5) to (v3);
\draw  [->] (q6) to (v5);
\draw  [->] (q6) to (v6);
\draw  [->, bend right] (q4) to (v5);
\draw  [->, bend left] (q5) to (v6);

\end{tikzpicture}
\caption{Two marked P/T nets that are fc-bisimilar, but not hfc-bisimilar}
\label{hfc-vs-fc}
\end{figure}
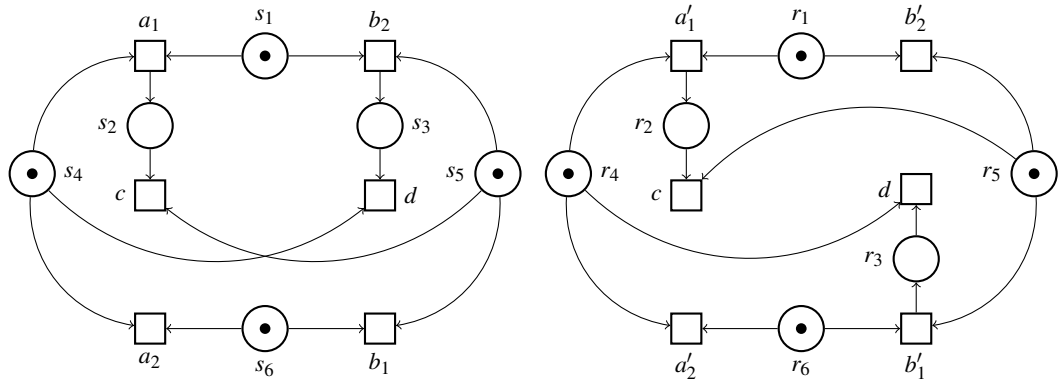

Finally, an example showing that hereditary
fully concurrent bisimilarity is finer than fully concurrent bisimilarity is outlined in Figure \ref{hfc-vs-fc}, originally
presented in \cite{NC94,FH99}, where labels $a_1, a_1', a_2, a_2'$ should be interpreted simply as $a$ 
and, similarly $b_1, b_1', b_2, b_2'$ simply as $b$.
Note that both systems have an $a$-action (a $b$-action) that can be followed by a dependent $c$-action ($d$-action) 
or an independent (because it is not competing on any places) $b$-action ($a$-action). 
Moreover, both have an $a$-action (a $b$-action) which can be followed by an independent $b$-action ($a$-action). 
Consequently, the two systems are fc-bisimilar. However, it is easy to argue that they are not hfc-bisimilar.
In fact, note that in order to match the causal sequence $a_1 c$,
it is necessary to match $a_1$ with $a_1'$, so that if, after that, $b_1$ is performed instead of $c$, then this can be only matched by $b_1'$, 
reaching the markings $s_2$ and $r_2 \oplus r_3$, respectively. 
But now, if we backtrack on $a_1$ and $a_1'$, then it is required that the markings $s_1 \oplus s_4$
and $r_1 \oplus r_3 \oplus r_4$ be hfc-bisimilar, but this is false, as only $r_1 \oplus r_3 \oplus r_4$ can perform $d$.

Not surprisingly, these two nets are not causal-net bisimilar. In fact,  whatever are the mappings from
the six conditions $\{b_1, b_2, b_3, b_4, b_5, b_6\}$ of the initial causal net to the initial markings of the two nets, we 
can always find a distinguishing transition that makes the two systems not equivalent. For instance, if
we map
$b_i$ to $s_i$ for the first net, and $b_i$ to $r_i$ for the second net for $i = 1, \ldots, 6$, then the 
transition $(r_5 \oplus r_6, b, r_3)$ of the second net, that enables a $d$-labeled transition, 
cannot be matched by $s_5 \oplus s_6$, because its $b$-labeled transition $b_1$ generates the empty marking $\theta$.
 
%
\section{Place Bisimilarity is Self-reversible} \label{place-reverse-sec}
%

Since $\sim_p$ is finer than $\sim_{cn}$, we expect that also place bisimilarity is self-reversible. 
In fact, consider two place bisimilar markings $m_1$ and $m_2$. For each transition $t_1$ performed by $m_1$, 
marking $m_2$ can respond with the transition $t_2$ that generates the same causal net, reaching a 
pair of place bisimilar markings; and this can be repeated from these reached markings. Of course, 
since the generated causal net is the same,
we are sure that also by undoing events, we end up into a pair of place bisimilar markings.

\begin{theorem}\label{place-rev-th}
Place bisimilarity is self-reversible.
\proof
Assume $m_1 \sim_p m_2$, i.e.,  a place bisimulation $R$ exists such that 
$(m_1, m_2) \in R^\oplus$. We will prove, by induction on the length of the computation from $m_1$, that there 
is a matching computation from $m_2$ such that the two computations are reversible. The case when $m_2$ moves first is symmetric,
and so omitted.

The base case is $n = 1$. If $m_1 [t_1\rangle m^1_1$, then, by Definition \ref{def-place-bis}, there exists $t_1'$ such that $m_2[t_1'\rangle m^1_2$ 
with $(\pre{t_1}, \pre{t_1'}) \in R^\oplus$, $\ell(t_1) = \ell(t_1')$,  $(\post{t_1}, \post{t_1'}) \in R^\oplus$ and, moreover, 
$(m_1^1, m_2^1) \in R^\oplus$. If we undo transitions $t_1$ and $t_1'$, respectively, we end up to the initial markings $m_1$ and $m_2$, 
which are place bisimilar by hypothesis.

Now assume $m_1[t_1\rangle m_1^1 [t_2\rangle m_1^2 \ldots [t_{n-1}\rangle m_1^{n-1} [t_n\rangle m_1^n$, with $n \geq 2$.
By induction, we know that there exists a computation
$m_2[t_1'\rangle m_2^1 [t_2'\rangle m_2^2 \ldots [t_{n-1}'\rangle m_2^{n-1}$ 
such that $(\pre{t_i}, \pre{t_i'}) \in R^\oplus$, $\ell(t_i) = \ell(t_i')$,  $(\post{t_i}, \post{t_i'}) \in R^\oplus$, $(m_1^i, m_2^i) \in R^\oplus$ 
 for $i = 1, \ldots, n-1$ and, moreover, that these computations of length $n-1$ are reversible.
We know that $(m_1^{n-1}, m_2^{n-1}) \in R^\oplus$ and that $m_1^{n-1} [t_n\rangle m_1^n$.
We have to distinguish two cases.

\begin{itemize}
\item
If $t_n$ does not consume a token produced by $t_{n-1}$, then the two transitions are independent.
Hence, $m_1^{n-1} = \post{t_{n-1}} \oplus \pre{t_n} \oplus m$ for some $m$. Since $(m_1^{n-1}, m_2^{n-1}) \in R^\oplus$, 
by Proposition \ref{add-prop1}(5), there exist
$\underline{m}^1_2, \underline{m}_2^2, m'$ such that $m_2^{n-1} = \underline{m}^1_2 \oplus \underline{m}_2^2 \oplus m'$
and $(\post{t_{n-1}}, \underline{m}^1_2)\in R^\oplus$, $(\pre{t_n}, \underline{m}^2_2)\in R^\oplus$
and $(m, m') \in R^\oplus$.

Since $(\pre{t_n}, \underline{m}^2_2)\in R^\oplus$, there exist a transition $t_n'$ such that
$\pre{t_n'} = \underline{m}^2_2$, $\ell(t_n) = \ell(t_n')$ and $(\post{t_n}, \post{t_n'}) \in R^\oplus$.
Therefore, $m_2^{n-1} [t_n'\rangle m_2^n$, where $(m^n_1, m^n_2) \in R^\oplus$, by additivity (Proposition \ref{add-prop1}(3)),
because $m^n_1 = \post{t_{n-1}} \oplus \post{t_n} \oplus m$ and $m^n_2 = \post{t_{n-1}'} \oplus \post{t_n'} \oplus m'$. 
Note that $t_{n-1}'$ and $t_n'$ are independent because $\underline{m}^1_2 = \post{t_{n-1}'}$.

If we backtrack on $t_n$ and $t_n'$, we end up in $(m^{n-1}_1, m^{n-1}_2) \in R^\oplus$, so this case is trivial.
If we backtrack on $t_{n-1}$ and $t_{n-1}'$, then we have
\begin{itemize}
\item 
$m_1^{n-2} [t_n\rangle \underline{m}^{n-1}_1 [t_{n-1} \rangle m^n_1$ and
\item 
$m_2^{n-2} [t_n'\rangle \underline{m}^{n-1}_2 [t_{n-1}' \rangle m^n_2$.
\end{itemize}
So, we have to prove that $(\underline{m}^{n-1}_1, \underline{m}^{n-1}_2) \in R^\oplus$; this follows easily by additivity because
$\underline{m}^{n-1}_1 = \pre{t_{n-1}} \oplus \post{t_n} \oplus m$ and $\underline{m}^{n-1}_2 = \pre{t_{n-1}'} \oplus \post{t_n'} \oplus m'$.

Induction can be invoked to conclude that also the computations of length $n - 1$ 

$m_1[t_1\rangle m_1^1 [t_2\rangle m_1^2 \ldots m_1^{n-2} [t_{n}\rangle \underline{m}^{n-1}_1$
and
$m_2[t_1'\rangle m_2^1 [t_2'\rangle m_2^2 \ldots m_2^{n-2} [t_{n}'\rangle \underline{m}^{n-1}_2$

\noindent
can match their past history transitions.
\item
If $t_n$ consumes a token produced by $t_{n-1}$, then the two transitions are causally dependent, so that we can 
backtrack on $t_n$ only. 
Since $(m_1^{n-1}, m_2^{n-1}) \in R^\oplus$, 
by Proposition \ref{add-prop1}(5), there exist
$\underline{m}^1_2, \underline{m}_2^2$ such that $m_2^{n-1} = \underline{m}^1_2 \oplus \underline{m}_2^2$, with 
$(\post{t_{n-1}}, \underline{m}^1_2) \in R^\oplus$ and $(m_1^{n-1} \ominus \post{t_{n-1}}, \underline{m}_2^2) \in R^\oplus$.

Now, consider the linking $l = 
(s_1, s_1') \oplus \ldots \oplus (s_k, s_k')$ that proves $(\post{t_{n-1}}, \post{t_{n-1}'}) \in R^\oplus$.
And consider the linking $h = (\overline{s}_1, \overline{s}_1') \oplus \ldots \oplus (\overline{s}_h, \overline{s}_h')$
that proves $(m_1^{n-1} \ominus \post{t_{n-1}}, \underline{m}_2^2) \in R^\oplus$.
We can extract from $l \oplus h$ a sublinking $r$ such that $\pi_1(r) = \pre{t_n}$ and $\pi_2(r) = \overline{m}$.
Note that $\overline{m} \cap \post{t_{n-1}'} \neq \emptyset$ by construction.
Hence, $m_1^{n-1} = m \oplus \pre{t_n}$ for some $m$, and $m_2^{n-1} = m' \oplus \overline{m}$ for some $m'$
such that $(m, m') \in R^\oplus$ (justified by the linking $l \oplus h \ominus r$) and $(\pre{t_n}, \overline{m}) \in R^\oplus$
(justified by the linking $r$). Therefore, there exist $t_n'$ such that
$\ell(t_n) = \ell(t_n')$, $\pre{t_n'} = \overline{m}$ and $(\post{t_n}, \post{t_n'}) \in R^\oplus$.
Hence, $m^{n-1}_2 [t_n' \rangle m^n_2$, where $(m^n_1, m^n_2) \in R^\oplus$, by additivity, as
$m^n_1 = m \oplus \post{t_n}$ and $m_2^n = m' \oplus \post{t_n'}$.

Note that $t_n'$ causally depends on $t_{n-1}'$, as it is consuming tokens produced by $t_{n-1}'$; hence, only $t_n'$ can be undone.
Therefore, by backtracking on $t_n$ and $t_n'$, respectively, we get the thesis $(m^{n-1}_1, m^{n-1}_2) \in R^\oplus$.
\end{itemize}
Summing up, we have shown that two place bisimilar markings $m_1$ and $m_2$
can match their forward behavior in such a way to ensure that also their backward 
behavior is preserved.
\fine
\end{theorem}

\section{A Small Case Study} \label{case-sec}
%

\begin{figure}[t]
\centering

\begin{tikzpicture}[
every place/.style={draw,thick,inner sep=0pt,minimum size=6mm},
every transition/.style={draw,thick,inner sep=0pt,minimum size=4mm},
bend angle=42,
pre/.style={<-,shorten <=1pt,>=stealth,semithick},
post/.style={->,shorten >=1pt,>=stealth,semithick}
]
\def\eofigdist{5cm}
\def\eodist{0.5cm}
\def\eodisty{1.5cm}

\node (p1) [place,tokens=1]  [label=above:$P_1$] {};
\node (p2) [place,tokens=1]  [right=\eodisty of p1,label=above:$C_1$] {};
\node (t1) [transition] [below=\eodist of p1,label=left:{\em prod}] {};
\node (p3) [place] [below =\eodist of t1,label=left:$D_1$] {};
\node (t2) [transition] [below right=\eodist of p3,label=left:{\em del}] {};
\node (p4) [place] [below right=\eodist of t2,label=left:$C_1'$] {};
\node (t4) [transition] [right=\eodist of p4,label=below:{\em cons}] {};

\draw  [->, bend left] (p1) to (t1);
\draw  [->, bend left] (t1) to (p1);
\draw  [->] (t1) to (p3);
\draw  [->] (p3) to (t2);
\draw  [->] (p2) to (t2);
\draw  [->] (t2) to (p4);
\draw  [->] (p4) to (t4);
\draw  [->] (t4) to (p2);

\node (q1) [place,tokens=1]  [right=\eofigdist of p1,label=above:$P_2$] {};
\node (q2) [place,tokens=1]  [right=\eodisty of q1,label=above:$C_2$] {};
\node (s1) [transition] [below=\eodist of q1,label=right:{\em prod}] {};
\node (q3) [place] [below left=\eodist of s1,label=above:$P_2'$] {};
\node (q'3) [place] [below right=\eodist of s1,label=left:$D_2'$] {};
\node (s2) [transition] [below left=\eodist of q3,label=left:{\em prod}] {};
\node (s'2) [transition] [below right=\eodist of q3,label=below:{\em prod}] {};

\node (q4) [place] [below right={1.2cm} of s2,label=left:$D_2''$] {};
\node (q5) [place] [below right={2.3cm} of s'2,label=below:$C_2'$] {};
\node (s3) [transition] [below right=\eodist of q4,label=left:{\em del}] {};
\node (s'3) [transition] [below right=\eodist of q'3,label=right:{\em del}] {};

\node (s4) [transition] [right=\eodist of q5,label=below:{\em cons}] {};

\draw  [->] (q1) to (s1);
\draw  [->] (s1) to (q3);
\draw  [->] (s1) to (q'3);
\draw  [->, bend left] (q3) to (s2);
\draw  [->, bend left] (s2) to (q3);
\draw  [->, bend left] (q3) to (s'2);
\draw  [->, bend left] (s'2) to (q3);

\draw  [->] (s2) to (q4);
\draw  [->] (s'2) to (q'3);

\draw  [->] (q4) to (s3);
\draw  [->] (q'3) to (s'3);
\draw  [->] (q2) to (s3);
\draw  [->] (q2) to (s'3);

\draw  [->] (s3) to (q5);
\draw  [->] (s'3) to (q5);
\draw  [->] (q5) to (s4);
\draw  [->, bend right] (s4) to (q2);

\end{tikzpicture}
\caption{The specification of an unbounded producer--consumer system (on the left) and one of its possible implementations (on the right)}
\label{upc1-place}
\end{figure}
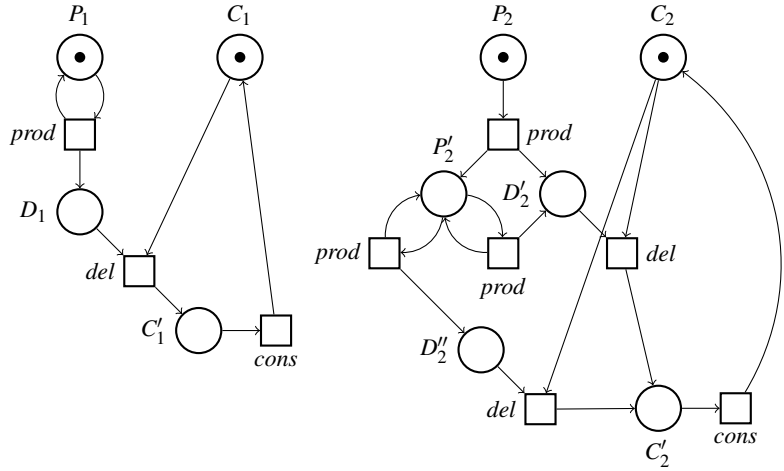

From the examples presented so far, it might seem that structure-preserving bisimilarity and place bisimilarity are 
too discriminating to be useful in practice. We want to contradict this opinion by means of the simple case study in Figure \ref{upc1-place},
which shows two forms of optimization which do not
alter the essential behavior of the system. We refer in
particular to replication (exemplified by transition {\em del}) and unfolding (exemplified by transition {\em prod}).

Consider the specification of an unbounded producer--consumer system and one of its possible implementations 
outlined in Figure \ref{upc1-place}, where {\em prod} is 
the action of producing an item,
{\em del} of delivering an item, {\em cons} of consuming an item. 
Note that if a token is present in the producer place $P_1$ (or $P_2'$), 
then the transition labeled by $prod$ can be performed at will, with the effect that the number of tokens 
accumulated into place $D_1$
(or $D_2'$, $D_2''$) can grow unboundedly; hence, the reachable markings from $P_1 \oplus C_1$ (or $P_2 \oplus C_2$)
are infinitely many, so that the state space of these two systems is infinite.

In order to prove that $P_1 \oplus C_1 \sim_p P_2 \oplus C_2$, we can try to build a place bisimulation $R$ which 
must contain at least the pairs
$(P_1, P_2), (C_1, C_2)$; then,
in order to match the $prod$-labeled transitions, we see that also the pairs $(P_1, P_2'),$ 
$ (D_1, D_2'), $ $(D_1, D_2'')$ are needed, and then,
in order to match the $del$-labeled transitions, also the pair $(C_1', C_2')$ must be considered. The resulting finite relation is 

$R = \{(P_1, P_2), (P_1, P_2'), (D_1, D_2'), (D_1, D_2''), (C_1, C_2), (C_1', C_2')\}$, 

\noindent which we will prove to be a place bisimulation, justifying that $P_1 \oplus C_1 \sim_p P_2 \oplus C_2$ because 
$(P_1 \oplus C_1, P_2 \oplus C_2) \in R^\oplus$. 

The proof that $R$ is a place bisimulation can be done by exploiting a characterization of place bisimulation
at the base of the proof of its decidability.
In fact, in \cite{Gor21decid,Gor25a} it is proved that a place relation $R \subseteq S \times S$ 
is a place bisimulation if and only if the following two finite conditions are satisfied:
\begin{enumerate}
\item $\forall t_1 \in T$, $\forall m$ such that $(\pre{t_1}, m) \in R^\oplus$
	\begin{itemize}
	\item[$(a)$] $\exists t_2$ such that $m = \pre{t_2}$, with  $\ell(t_1) = \ell(t_2)$ and $(\post{t_1}, \post{t_2}) \in R^\oplus$. 
	\end{itemize}
\item $\forall t_2 \in T$, $\forall m$ such that $(m, \pre{t_2}) \in R^\oplus$
	\begin{itemize}
	\item[$(b)$] $\exists t_1$ such that $m = \pre{t_1}$,
	with $\ell(t_1) = \ell(t_2)$ and $(\post{t_1}, \post{t_2}) \in R^\oplus$.
	\end{itemize}
\end{enumerate}

Hence, to check that $R$ is a place bisimulation, it is sufficient to make the checks above for each 
transition in the two nets. There are nine transitions, namely:\\
$\begin{array}{rcl@{\quad}|rclrcl}
t_1 &  =  & (P_1, prod, P_1 \oplus D_1) &\quad t_4 &  =  & (P_2, prod, P_2' \oplus D_2') & \quad t_7 & = & (D_2' \oplus C_2, del, C_2') \\
t_2 & = & (D_1 \oplus C_1, del, C_1')  &\quad t_5 &  =  & (P_2', prod, P_2' \oplus D_2'')& \quad t_8 & = & (D_2'' \oplus C_2, del, C_2') \\
t_3 & = & (C_1', cons, C_1) & \quad t_6 & = & (P_2', prod, P_2' \oplus D_2') & \quad t_9 & = & (C_2', cons, C_2) \\
\end{array}$\\
For each transition $t_i$ for $i = 1, \ldots, 3$, we have to find all the markings $m$ such that $(\pre{t_i}, m) \in R^\oplus$ and for each of them, to find
a matching $t_j$ (with $j = 4, \ldots, 9$) such that $\pre{t_j} = m$ and $(\post{t_i}, \post{t_j}) \in R^\oplus$. Symmetrically, for each transition $t_j$, we have to find all the markings $m$ such that $(m, \pre{t_j}) \in R^\oplus$ and for each of them,
to find a matching $t_i$ such that $\pre{t_i} = m$ and $(\post{t_i}, \post{t_j}) \in R^\oplus$.
For instance,  consider $t_2$, whose pre-set is $D_1 \oplus C_1$; there are two markings related to $D_1 \oplus C_1$,
namely $D_2' \oplus C_2$ and $D_2'' \oplus C_2$; in the former case, the matching transition is $t_7$, in the latter case $t_8$.
The reader can easily check that this is the case for all the transitions, so that we can conclude that $R$
is a place bisimulation, indeed.

Since $R$ is a place bisimulation, we have that $\mathcal{M}(R)$ is a structure-preserving bisimulation
proving that $P_1 \oplus C_1 \sim_{sp} P_2 \oplus C_2$.

%
\section{Conclusion and Future Research}\label{conc-sec}
%
We tackled the question, not previously considered in
the literature, of whether truly-concurrent bisimilarities for Petri
nets are self-reversible, meaning that they require to match
behaviors also going backwards. As main
results we have shown that causal-net bisimilarity is self-reversible,
and that place bisimilarity is reversible. The latter is of particular
interest since place bisimilarity is decidable for finite (but
possibly unbounded) Petri nets, what enables us to use it to prove the correctness 
of non-trivial implementations of given specifications, as we did in Section~\ref{case-sec}.

An obvious question is whether causal-net bisimilarity is the coarsest self-reversible behavioral relation fully respecting causality.
The answer to this question is negative. In fact, it is possible to prove, by a simple adaptation of the proof of Theorem \ref{cn=hcn-th},
that the coarser {\em i-causal-net bisimilarity} \cite{CG21a} is also self-reversible. This equivalence, which 
ensures that related markings generate 
the same causal nets, lies in between causal-net bisimilarity and hhp-bisimilarity; in fact, the markings in 
Figure \ref{cn-vs-icn-fig} are i-causal-net bisimilar (but not causal-net bisimilar), 
while the two nets in Figure 
\ref{net-fc-place} are hereditary history-preserving bisimilar (but not i-causal-net bisimilar). 
I-causal-net bisimilarity is decidable on bounded Petri nets \cite{CG21a}; this proof is based on a rather intricate
equivalent bisimulation-based characterization over {\em ordered indexed marking} (markings whose tokens are indexed, 
to give identity to multiple tokens on a place, and onto which a generation ordering is defined), and this explains why 
we have chosen to work directly with
causal-net bisimilarity, which is characterized in a very simple manner by structure-preserving bisimilarity.
We conjecture that i-causal-net bisimilarity is the coarsest self-reversible causality-respecting bisimulation-based behavioral 
relation definable on Petri nets.

The problem we tackled differs from the ones in \cite{B+18,ML19} or \cite{MMU20,dKM21},where the focus is respectively to understand the impact of adding the reverse of a single transition to a Petri net, or to add reverses of every transition so 
to be able 
to go backward on the history.
On the contrary, we do not
add reverse transitions to
the Petri net, but rather we define a {\em reversible behavioral equivalence} that ensures that behaviorally equivalent
markings can match not only their forward behavior, but also their backward behavior by undoing events.
We conjecture that if two nets are equivalent according to a reversible behavioral equivalence, their extensions as defined in \cite{MMU20} 
are equivalent as well.

Another direction for future work is to apply our results to CCS.
  Indeed, CCS and other process algebras have been equipped with Petri net semantics 
(see, e.g., \cite{DDM88,Old,Gor17}, just to mention a few). In \cite{Gor25a} one of the authors shows that a dialect of CCS, 
called {\bf Pet}, is sufficiently expressive to represent all and only the finite P/T nets.
For instance, the specification producer--consumer system in Figure \ref{upc1-place} is represented as $\restr{a}(P_1 \para C_1)$, where

$
\begin{array}{rclrclrclrclrcl}
 P_1 & \eqdef & {\mbox{\it prod}}.(P_1 \para D_1) & \quad & D_1 & \eqdef & \overline{a}.\1 & \quad & C_1 & \eqdef & \atom{a\,del}.C_1' & \quad &
 C_1' & \eqdef & {\mbox{\it cons}}.C_1\\
\end{array}$

\noindent $\1$ denotes successful termination, and $\atom{a ,  del}$ denotes that the sequence $a ,  del$ is to be performed atomically.
Hence, thanks to the results proved here, we conclude that 
place bisimilar {\bf Pet} process terms are able to match their forward and backward behavior in a causal-consistent 
manner. Moreover, place bisimilarity is a congruence over {\bf Pet}, so that the verification that an implementation is equivalent to its specification can be done compositionally.


\end{document}